\documentclass[aps,prd,twocolumn]{revtex4-2}

\usepackage[colorlinks]{hyperref}
\usepackage{amssymb,amsmath}
\usepackage{graphicx}

\begin{document}
\title{\bf 
Limits on scalar dark matter interactions with particles other than the photon \\ 
via loop corrections to the scalar-photon coupling
}
\author{V. V. Flambaum}\email{v.flambaum@unsw.edu.au}
\author{I. B. Samsonov}\email{igor.samsonov@unsw.edu.au}
\address{School of Physics, University of New South Wales,
Sydney 2052, Australia}

\begin{abstract}
There is limited information about the interaction strength of a scalar dark matter candidate with hadrons and leptons for a scalar particle mass exceeding $10^{-3}$\,eV while its interaction with photon is well studied. The scalar-photon coupling constant receives quantum corrections from one-loop Feynman diagrams which involve the scalar-lepton, scalar-quark, and scalar-W boson vertices. We calculate these one-loop quantum corrections and find new limits on the scalar particle interactions with electron, muon, tau, quarks, nucleons, gluons, Higgs, and W bosons by re-purposing the results of experiments measuring the scalar-photon interaction. Limits on interactions of heavy leptons and quarks have been obtained for the first time, and limits on other interactions in certain mass intervals are 2 to 15 orders of magnitude stronger than those presented in previous publications and exclude the resolution of the muon $g-2$ anomaly with scalar particle.
\end{abstract}

\maketitle

\section{Introduction}
Scalar and pseudoscalar fields are very promising candidates for dark matter (DM) particles. The pseudoscalar DM candidates are exemplified by the QCD axion which represents a consistent extension of the Standard Model and helps in resolving the strong CP problem (see, e.g., \cite{SnowmassAxion} for a review). The CP-even counterpart of the axion field is a dilaton $\phi$ \cite{dilaton}, which is strongly motivated by superstring theory, see, e.g., Refs.~\cite{dilaton1,dilaton2,dilaton3,dilaton4,dilaton5,dilaton6}. Scalar field DM candidates also appear in chameleon models of gravity, see, e.g., Ref.~\cite{chameleon}.

Since the preferred interaction channel between the scalar field and the visible matter particles is not known, it is generally assumed that all types of interactions are allowed, and the strengths of these interactions should be measured experimentally. The leading interaction vertices should be linear with respect to the scalar field, although quadratic interactions are also of interest, see, e.g., Refs.~\cite{2phi1,Stadnik2015,Stadnik}. Therefore, the majority of experiments searching for the scalar DM candidates focus on measurements of coupling constants of this scalar field to photon, $g_{\phi\gamma}$, and to electron, $g_{\phi e}$. The results of these measurements are summarized in Refs.~\cite{SnowmassScalar,AxionLimits}. The channel of hadronic interactions of this scalar field remains less studied.

In this paper, we re-purpose the results of the experiments measuring the scalar-photon interaction constant for obtaining new limits on the strength of the interaction of the scalar field with leptons,  quarks, gluons, Higgs and W bosons. The main idea is that the scalar-photon coupling constant $g_{\phi\gamma}$ receives quantum corrections from one-loop Feynman diagrams involving the scalar-fermion and scalar-W boson vertices. In other words, the scalar field can decay into two photons via fermionic and W boson loops. Below, we present the general expression for quantum corrections to the scalar-photon interaction constant from one-loop Feynman graphs. This process is analogous to the decay of the Higgs field into two photons mediated by leptons, W bosons or heavy quarks calculated in Refs.~\cite{Ellis1975,Shifman1979} (for light quarks the meson mechanism is more adequate). Then, we apply the one-loop relation between the coupling constants for obtaining new limits on the interactions of the scalar DM candidate.

We use natural units with $\hbar=c=1$.

\section{One-loop contribution to scalar-photon coupling}
In this section, we summarize known results for one-loop quantum contributions to scalar-photon coupling which will be used in the subsequent section.

\begin{figure}
    \includegraphics*[width=8cm]{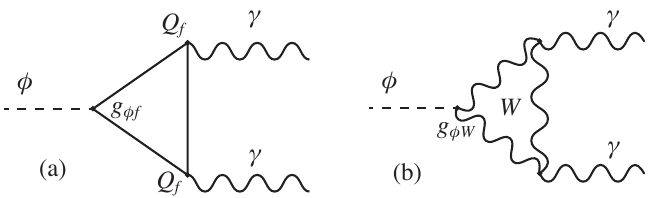}
    \caption{Feynman graph representing leading one-loop quantum correction to the scalar-photon coupling. (a) Contribution from fermion (lepton or heavy quark) fields with charges $Q_f$. (b) Contribution from W boson loop.}
    \label{fig1}
\end{figure}

Let $\phi$ be a real scalar field representing DM particles with mass $m_\phi$. In the model of dilaton-like DM \cite{Damour,Arvanitaki,Stadnik,Stadnik2015}, this field may have the following interaction vertices to the photon field with Maxwell field strength $F_{\mu\nu}$, to quarks or leptons described by Dirac fermions $f$, ($f=e,\mu,\tau,u,d,s,c,b,t$ for electron, muon, tau, up, down, strange, charm, bottom and top quarks, respectively), to gluons $G^a_{\mu\nu}$ and to $W^\pm_\mu$ bosons
\begin{equation}
\begin{aligned}
    {\cal L}_\text{int} &= \frac14 g_{\phi\gamma} \phi F^{\mu\nu}F_{\mu\nu} - \sum_{f}g_{\phi f} \phi \bar ff \\
    &+\frac14 g_{\phi g}\phi G^{\mu\nu a}G_{\mu\nu}^a+g_{\phi W}m_W \phi W^{+\mu}W^-_\mu\,.
\end{aligned}
\label{Lint}
\end{equation}
The coupling constant $g_{\phi\gamma}$  receives quantum corrections from fermionic loop diagrams with the scalar-fermion interaction quantified by the coupling constant $g_{\phi f}$ and from the W-boson loop involving the vertex with $g_{\phi W}$ coupling. The leading contributions are represented by the triangle Feynman graphs in Fig.~\ref{fig1}. The corresponding scattering amplitude ${\cal M}$ reads as
\begin{equation}
\begin{aligned}
    {\cal M} &= \frac{\alpha}{2\pi}\left(\frac{g_{\phi f}}{m_f} A_{1/2}+\frac{g_{\phi W}}{2m_W}A_1\right) 
    \\  &\times(k_1^\rho k_{2\rho} g^{\mu\nu} - k_1^\mu k_2^\nu) e_{1\mu}e_{2\nu}\,,
\end{aligned}
\label{M}
\end{equation}
where $m_f$ and $m_W$ are the fermion particle and W boson rest masses, respectively, $k_1^\mu$, $k_2^\nu$ and $e_{1\mu}$, $e_{2\nu}$ are four-momenta and polarization four-vectors of the photons. The dimensionless functions $A_{1/2}$ and $A_1$ were calculated in Refs.~\cite{Ellis1975,Shifman1979} where decay of the Higgs boson into two photons was considered (see also Ref.~\cite{HiggsReview} for a review):
\begin{subequations}
\label{Afunctions}
\begin{align}
    A_{1/2} &= 2[\tau +(\tau-1)f(\tau)]\tau^{-2}\,,\\
    A_1 &= -[2\tau^2 + 3\tau +3(2\tau-1)f(\tau)]\tau^{-2}\,,
\end{align}
\end{subequations}
where 
\begin{equation}
    f(\tau) = \left\{ 
    \begin{array}{ll}
    \arcsin^2\sqrt\tau & \tau\leq1 \\
    -\frac14 \left[
    \log\frac{1+\sqrt{1-\tau^{-1}}}{1-\sqrt{1-\tau^{-1}}} -i\pi
    \right]^2 &\tau>1\,,
    \end{array}
    \right. 
\end{equation}
and $\tau \equiv m_\phi^2/(4m_f^2)$ for fermions with mass $m_f$ or $\tau \equiv m_\phi^2/(4m_W^2)$ for $W$-boson with mass $m_W$.

For our estimates, asymptotic values of functions (\ref{Afunctions}) at small arguments will suffice,
\begin{subequations}
\begin{align}
    A_{1/2} &\approx \frac43 \sum_f N_c Q_f^2 && (m_\phi\ll 2m_f),\label{Alight}\\
    A_1 &\approx -7 &&(m_\phi\ll 2m_W).
    \label{Wheavy}
\end{align}
\end{subequations}
Here $N_c$ is the number of colors ($N_c=3$ for quarks and $N_c=1$ for leptons) and $Q_f$ are fractional quark and lepton charges ($Q_f=+\frac32$ for $c$ and $t$ quarks, $Q_f=-\frac13$ for $b$ quarks, and $Q_f=-1$ for electron, muon, and tau). 

The expression for the scattering amplitude (\ref{M}) provides the one-loop contribution to the scalar-photon coupling constant $g_{\phi\gamma}$:
\begin{equation}
    g_{\phi\gamma} = \frac{\alpha}{2\pi} \left( \frac{g_{\phi f}}{m_f} A_{1/2} +\frac{g_{\phi W}}{2m_W} A_1 \right)\,.
\label{result}
\end{equation}
Below, we will use this relation for extracting limits on $g_{\phi f}$ and $g_{\phi W}$ from known experimental constraints on the scalar-photon coupling $g_{\phi\gamma}$.

\section{Limits on the interaction with standard model particles}
\subsection{Scalar-electron interaction}
\begin{figure}
    \centering
    \includegraphics[width=8cm]{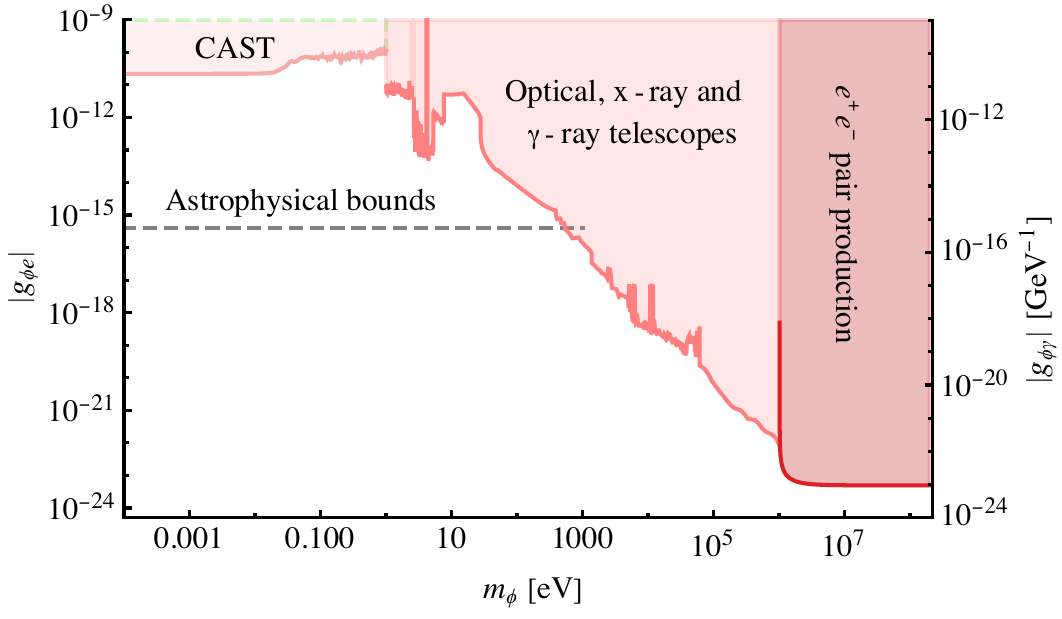}
    \caption{Limits on the scalar-photon $g_{\phi\gamma}$ (right scale) and scalar-electron $g_{\phi e}$ (left scale) interaction constants. Light pink excluded area originates from re-purposing the results of the CAST experiment \cite{CAST}. Pink area represents the consolidated excluded region arising from application of the relation (\ref{g-lignt}) to the results of the works \cite{XRAY,XMM,NuStar1,NuStar2,NuStar3,INTEGRAL,LeoT,GRayAtt,HST,HSTNak,VIMOS,CMB1,CMB2,MUSE1,MUSE2,JWST,ROSITA}. Dark pink area sets an upper limit on the scalar-electron coupling constant $g_{\phi e}$ from the direct tree-level decay of scalars into electron-positron pairs. Gray dashed line shows recent astrophysical bounds on $g_{\phi e}$ from Ref.~\cite{Vitagliano}. Limits on the scalar-gluon interaction constant $g_{\phi g}$ are about 3 orders in magnitude weaker than limits on $g_{\phi\gamma}$.}
    \label{fig:electron}
\end{figure}

Since the mass of the scalar field $\phi$ is unknown, it is necessary to search for a scalar DM signal in a wide range of masses, spanning from ultralight fields with masses $m_\phi\sim 10^{-24}$\,eV to the MeV-GeV region. If the scalar field is much lighter than the electron, $m_\phi\ll m_e$, the identity \eqref{Alight} applies with $N_c = 1$ and $Q_f^2 = 1$. In this case, the one-loop contribution to the scalar-photon coupling constant from the electron loop \eqref{result} reads as
\begin{equation}
g_{\phi\gamma} = \frac{2\alpha}{3\pi m_e} g_{\phi e}\,.
\label{g-lignt}
\end{equation}
This relation allows us to find the limits on 
$g_{\phi e}$ from the known constraints on the coupling constant $g_{\phi\gamma}$ 
in the mass region $10^{-24}\,\text{eV}<m_\phi\lesssim 10^{6} \,\text{eV}$. Limits on $g_{\phi\gamma}$ for mass $m_\phi\lesssim 10^{-4} \,\text{eV}$ are presented  in Refs.~\cite{SnowmassScalar,AxionLimits,phi-gamma1,phi-gamma2,phi-gamma3,phi-gamma4,phi-gamma5,phi-gamma6,phi-gamma7,phi-gamma8,phi-gamma9,phi-gamma10,phi-gamma11,phi-gamma12,phi-gamma13,phi-gamma14,phi-gamma15,phi-gamma16,phi-gamma17,phi-gamma18,phi-gamma19,my1,my2}. 
However, in the interval $10^{-24}\,\text{eV}<m_\phi\lesssim 10^{-4}\,\text{eV}$ our limits for $g_{\phi e}$ are 1--2 orders in magnitude weaker than the direct limits on the scalar-electron interaction obtained in the works \cite{phi-gamma4,phi-gamma5,phi-gamma8,phi-gamma9,phi-gamma10,phi-gamma11,phi-gamma12,phi-gamma14,phi-gamma15,phi-gamma16,phi-gamma17,phi-gamma18,phi-gamma19,AURIGA,Cavities,YbCs,NANOGrav} through direct measurements of this coupling. Therefore, we only present limits for $m_\phi > 10^{-4}\,\text{eV}$ in Fig.~\ref{fig:electron}.

Scalar particles with mass in the eV to sub-MeV range can decay into photons and create emission lines potentially observable with the aid of space-based optical, x-ray, and gamma-ray telescopes. The decay rate of these scalars is the same as that for the axion-like particles with $g_{a\gamma}$ replaced by $ g_{\phi\gamma}$. Therefore, making use of the identity \eqref{g-lignt} we re-purpose the results of the works \cite{AxionLimits,XRAY,XMM,NuStar1,NuStar2,NuStar3,INTEGRAL,LeoT,GRayAtt,HST,HSTNak,VIMOS,CMB1,CMB2,MUSE1,MUSE2,JWST,ROSITA} to obtain the limits on the scalar-photon $g_{\phi\gamma}$ and scalar-electron $g_{\phi e}$ coupling constants. These limits are shown in Fig.~\ref{fig:electron} as a consolidated pink area in the range of scalar filed masses $0.8\,\text{eV}<m_\phi<1\,\text{MeV}$. 

Figure \ref{fig:electron} includes also limits on $g_{\phi\gamma}$ and $g_{\phi e}$ in the interval $10^{-4}\,\text{eV}<m_\phi<1\,\text{eV}$ which originate from the limits of the CAST experiment \cite{CAST} on the axion-photon interaction $g_{a\gamma}$. In Ref.~\cite{Tobar}, it was shown that the results of this experiment apply to the scalar-photon interaction with $g_{a\gamma}\to g_{\phi\gamma}$. Thus, making use of Eq.~(\ref{g-lignt}), we re-purpose the results of the CAST experiment \cite{CAST} to obtain the limits on the scalar-electron interaction constant $g_{\phi e}$. For a comparison, we included in Fig.~\ref{fig:electron} recent astrophysical bounds on $g_{\phi e}$ from Ref.~\cite{Vitagliano} (gray dashed line).
 
If the mass of the scalar field $m_\phi>2m_e$, the scalar can decay directly into electron-positron pairs. The tree-level decay rate of this process is \cite{MuonPospelov} (see also Ref.~\cite{HiggsReview})
\begin{equation}
    \Gamma(\phi\to e^+e^-) = \frac1{8\pi} g_{\phi e}^2 m_\phi \left(1-\frac{4m_e^2}{m_\phi^2}\right)^{3/2}.
\label{Gamma}
\end{equation}
The electrons and positrons ejected in the decays of scalar DM particles serve as an additional source of diffuse photons produced by the interaction with an interstellar medium. The ejected positrons,  after loosing their kinetic energy, annihilate in the interstellar medium and produce the 511 keV photon line in gamma-ray detectors such as the SPI/INTEGRAL telescope. The reported flux of 511 keV photons from the Galactic bulge is 
$\Phi \simeq 10^{-3} \text{cm}^{-2}\text{s}^{-1}$ \cite{SPI-INTEGRAL}. Assuming that a fraction of this photon flux is  produced by decays of the scalar dark matter particles, we can set an upper bound on the value of the coupling $g_{\phi e}$ in Eq.~(\ref{Gamma}). The corresponding limits on $g_{\phi e}$ are represented by a dark pink area in Fig.~\ref{fig:electron} 
 in the mass region $2m_e<m_\phi \lesssim  210 $\,MeV (for $m_\phi>2m_\mu$ other effects such as production of pairs of mesons may dominate). In calculating these limits, we followed the procedure similar to the one in Ref.~\cite{QN1}.

\begin{figure*}
    \centering
    \begin{tabular}{cc}
    (a) & (b) \\
    \includegraphics[width=8cm]{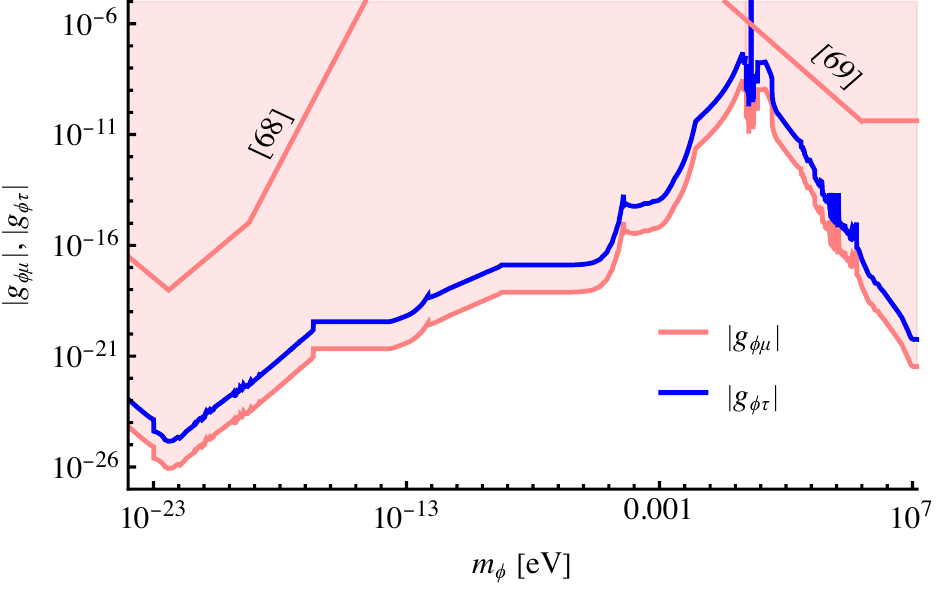} &
    \includegraphics[width=8.8cm]{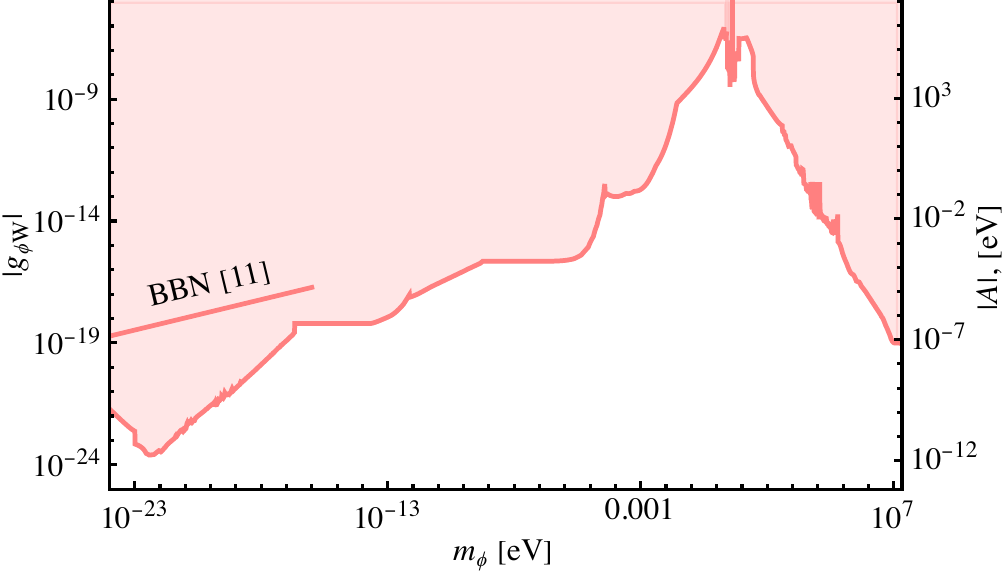}
    \end{tabular}
    \caption{(a) Limits on scalar-muon (pink curve) and scalar-tau (blue curve) coupling constants obtained by applying the relation (\ref{mutau}) to the results of the works \cite{phi-gamma1,phi-gamma2,phi-gamma3,phi-gamma4,phi-gamma5,phi-gamma6,phi-gamma7,phi-gamma8,phi-gamma9,phi-gamma10,phi-gamma11,phi-gamma12,phi-gamma13,phi-gamma14,phi-gamma15,phi-gamma16,phi-gamma17,phi-gamma18,phi-gamma19} and \cite{XRAY,XMM,NuStar1,NuStar2,NuStar3,INTEGRAL,LeoT,GRayAtt,HST,HSTNak,VIMOS,CMB1,CMB2,MUSE1,MUSE2,JWST}. (b) Analogous limits on the scalar-W boson ($g_{\phi W}$, left scale) and scalar-Higgs ($A$, right scale) coupling constants.
    }
    \label{fig:leptonsW}
\end{figure*}

\subsection{Scalar-muon and tau interactions}
Equation (\ref{g-lignt}) may be easily generalized to accommodate one-loop contributions from the muon and tau loops,
\begin{equation}
g_{\phi\gamma} = \frac{2\alpha}{3\pi} \left( \frac{g_{\phi\mu}}{m_\mu} + \frac{g_{\phi\tau}}{m_\tau}\right)\,,
\label{mutau}
\end{equation}
where $m_\mu$ and $m_\tau$ are muon and tau rest masses, respectively. Note that these masses are significantly larger than the upper bound on the scalar field mass considered in the present study, $m_\phi<15$\,MeV. Therefore, we do not need to consider direct decay of the scalar into muon and tau pairs as in Eq.~(\ref{Gamma}). In Fig.~\ref{fig:leptonsW}a we present the limits on the coupling constants $g_{\phi\mu}$ and $g_{\phi\tau}$ obtained by applying Eq.~(\ref{mutau}) to the results of the works \cite{phi-gamma1,phi-gamma2,phi-gamma3,phi-gamma4,phi-gamma5,phi-gamma6,phi-gamma7,phi-gamma8,phi-gamma9,phi-gamma10,phi-gamma11,phi-gamma12,phi-gamma13,phi-gamma14,phi-gamma15,phi-gamma16,phi-gamma17,phi-gamma18,phi-gamma19,my1,my2} and \cite{CAST,XMM,NuStar1,NuStar2,NuStar3,INTEGRAL,LeoT,GRayAtt,HST,HSTNak,VIMOS,CMB1,CMB2,MUSE1,MUSE2,JWST,ROSITA}. Our limits on $g_{\phi\mu}$ exclude the resolution of the muon $g-2$ anomaly with scalar particle \cite{MuonPospelov} for $m_\phi<15$\,MeV. 

Note that our limits on the scalar-muon coupling represent an 8--15 orders in magnitude improvement over the results of the works \cite{StadnikMuon,MuonPospelov} and 2--11 orders in magnitude improvement over the recent astrophysical bounds \cite{Raffelt}. Limits on the scalar-tau interaction are presented for the first time to the best of our knowledge.

\subsection{Scalar-W boson interaction}
Let us now assume that the dominant one-loop contribution to the scalar-photon coupling arises from the loop of W bosons shown in Fig.~\ref{fig1}b. Noting that in the case under consideration the W-boson mass is always much larger than the scalar field mass, Eqs.~(\ref{Wheavy}) and (\ref{result}) may be cast in the form
\begin{equation}
    g_{\phi\gamma} = -\frac{7\alpha}{4\pi m_W} g_{\phi W}\,.
\label{21}
\end{equation}
This relation allows us to find limits on $g_{\phi W}$ from known constraints on the scalar-photon interaction presented in Refs.~\cite{phi-gamma1,phi-gamma2,phi-gamma3,phi-gamma4,phi-gamma5,phi-gamma6,phi-gamma7,phi-gamma8,phi-gamma9,phi-gamma10,phi-gamma11,phi-gamma12,phi-gamma13,phi-gamma14,phi-gamma15,phi-gamma16,phi-gamma17,phi-gamma18,phi-gamma19,XRAY,XMM,NuStar1,NuStar2,NuStar3,INTEGRAL,LeoT,GRayAtt,HST,HSTNak,VIMOS,CMB1,CMB2,MUSE1,MUSE2,JWST}, see Fig.~\ref{fig:leptonsW}b. Our limits on $g_{\phi W}$ are 2--5 orders in magnitude stronger than those found earlier in Ref.~\cite{Stadnik2015} for $m_\phi<10^{-16}$\,eV from the analysis of big bang nucleosynthesis.

\subsection{Scalar-quark interaction}
One-loop contributions to $g_{\phi\gamma}$ from ``light'' $u,d,s$ and ``heavy'' $c,b,t$ quarks should be considered slightly differently. Let us start with the light quarks first.

\begin{figure}
    \includegraphics*[width=8cm]{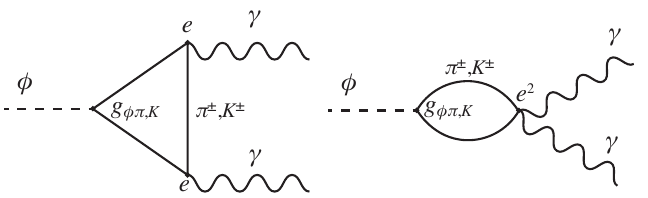}
    \caption{Feynman graphs representing leading one-loop quantum correction to the scalar-photon coupling from charged mesons.}
    \label{figMesonLoops}
\end{figure}

\begin{figure}
    \centering
    \includegraphics[width=8cm]{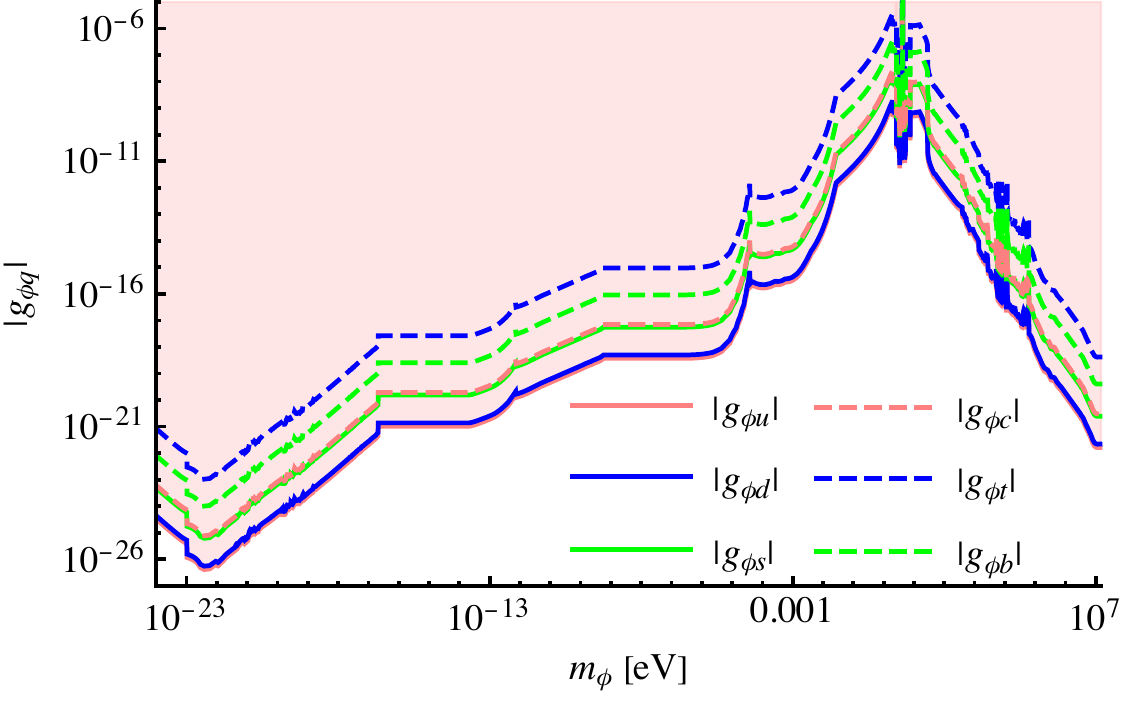}
    \caption{Limits on scalar-quark interaction.}
    \label{fig:quarks}
\end{figure}

For the contribution from the $u$, $d$ and $s$ quark loops, the relations (\ref{Alight}) and (\ref{result}) cannot be applied naively, because they do not take into account the effect of strong coupling of quarks at low energies. Indeed, below the QCD scale, the QCD theory reduces to the effective theory of mesons known as the chiral perturbation theory. In this theory, the leading contributions to the scalar decay rate to mesons are represented by Feynman diagrams in Fig.~\ref{figMesonLoops}. These contributions were calculated in Ref.~\cite{Leutwyler89}. In the Appendix \ref{AppA}, we give some details of these calculations and derive the contributions to the scalar-photon interaction constant from the scalar-quark ones:
\begin{equation}
    g_{\phi\gamma} \approx [0.030 g_{\phi u} + 0.028 g_{\phi d} + 0.0020 g_{\phi s}]\text{GeV}^{-1}\,.
\label{16}
\end{equation}
This allows us to derive limits on either of the constants $g_{\phi u}$, $g_{\phi d}$ and $g_{\phi s}$ assuming that it gives a dominant contribution to $g_{\phi\gamma}$ (see Fig.~\ref{fig:quarks}). The same experimental searches as for the analysis of lepton couplings have been used.

For ultralight scalars with $m_\phi \lesssim 0.05$\,eV our limits on the light quarks are 3--5 orders in magnitude weaker than the existing ones obtained directly from the torsion-balance and equivalence principle test experiments \cite{EP1,EP2,EP3,EP4,EP5,EP6,Leefer2016,phi-gamma4,phi-gamma6}, as well as from an analysis of atomic clock frequency shifts
\cite{phi-gamma1,phi-gamma2,phi-gamma3,phi-gamma6,phi-gamma7,phi-gamma8}.

For the one-loop contributions to $g_{\phi\gamma}$ from heavy quarks $c$, $b$, and $t$ the effects of strong interaction are not as important as for the light ones. Therefore, bare masses of these quarks may be used in Eq.~(\ref{result}),
\begin{equation}
    g_{\phi\gamma} = \frac{2\alpha}{\pi} \left[
        \frac49 \frac{g_{\phi c}}{m_c} + \frac19 \frac{g_{\phi b}}{m_b} + \frac49 \frac{g_{\phi t}}{m_t} \right]\,.
\label{160}
\end{equation}
This relation allows us to find limits on $g_{\phi c}$, $g_{\phi b}$, and $g_{\phi t}$ from the constraints on $g_{\phi\gamma}$ assuming that either of these interactions is dominant. These limits are represented in Fig.~\ref{fig:quarks} by dashed curves.

\subsection{Order-of-magnitude estimate of scalar-nucleon interaction}
The couplings $g_{\phi u}$ and $g_{\phi d}$ may be related with the scalar-proton $g_{\phi p}$ and scalar-neutron $g_{\phi n}$ ones,
\begin{equation}
    g_{\phi p} = 2 g_{\phi u} + g_{\phi d}\,,\qquad
    g_{\phi n} = g_{\phi u} + 2 g_{\phi d}\,.
\end{equation}
Substituting these relations into Eq.~(\ref{16}) and ignoring contributions from $g_{\phi s}$, we find
\begin{equation}
    g_{\phi\gamma} = (8.6 g_{\phi p} - 1.2 g_{\phi n})10^{-3}\, \text{GeV}^{-1}\,.
\end{equation}
This equation allows us to set limits on the scalar-nucleon interaction constants $g_{\phi p}$ and $g_{\phi n}$ assuming that either of these contributions is dominant. These limits are very close to the ones for the scalar-quark interaction in Fig.~\ref{fig:quarks}. Therefore, we do not present here dedicated plots for the scalar-nucleon constants, but we point out that these limits are about 4 orders in magnitude stronger than the ones obtained in Ref.~\cite{Dzuba23} from the analysis of results of liquid noble gas DM search experiments but 1--3 orders in magnitude weaker than the recent astrophysical bounds from Ref.~\cite{Vitagliano} for $m_\phi\lesssim 200$~eV.

\subsection{Scalar-Higgs interaction}
The interaction of the scalar field $\phi$ with the Higgs doublet $H$ was considered in Ref.~\cite{PospelovHiggs}, ${\cal L}_\text{int} = -A \phi H^\dagger H$, where $A$ is a coupling constant of dimension energy. There  was shown that this interaction gives the following one-loop quantum contribution to the scalar-photon coupling constant:
\begin{equation}
    g_{\phi\gamma} = \frac{\alpha}{2\pi m_h^2} A\approx 7.4\times 10^{-8}\text{\,GeV}^{-2}A\,,
\end{equation}
where $m_h=125$\,GeV is the Higgs mass. This relation allows us to find limits on the scalar-Higgs field interaction from the constraints on the scalar-photon interaction given above. These limits on the coupling constant $A$ extend  the results of work \cite{StadnikHiggs} to heavier scalars, see Fig.~\ref{fig:leptonsW}b.

\subsection{Order-of-magnitude estimate of scalar-gluon interaction}
The contribution of the scalar-gluon vertex in Eq.~(\ref{Lint}) to the scalar-photon interaction constant starts from two loops. This contribution may be roughly estimated as
\begin{equation}
    g_{\phi\gamma} = K \frac{\alpha}\pi \frac{\alpha_s}\pi  g_{\phi g}\,,
\label{gstrong}
\end{equation}
where $K$ is a $O(1)$ numeric coefficient and $\alpha_s\sim1$ is the strong interaction constant. Equation (\ref{gstrong}) gives an order-of-magnitude estimate for the scalar-gluon coupling constant from the experimentally measured value of $g_{\phi\gamma}$. An accurate calculation of the coefficient $K$ in Eq.~(\ref{gstrong}) goes beyond the scope of this paper since it involves non-perturbative aspects of QCD at low energy.

\section{Summary}

In this work, we have derived new limits on the strength of interaction of the hypothetical scalar dark matter particle with various Standard Model particles and for scalar field masses in the interval $10^{-23}\,\text{eV} < m_\phi < 15\,\text{MeV}$. These limits are found by re-purposing the known limits on the scalar-photon and axion-photon coupling constants from various terrestrial experiments and cosmological observations. For deriving these limits we employed Eq.~(\ref{result}) which represents one-loop quantum contributions to the scalar-photon coupling from fermions and W-boson.

Our limits on the scalar-muon coupling are significantly stronger, by 8--15 orders of magnitude, than those reported in Refs. \cite{MuonPospelov,StadnikMuon}, and 2--11 orders of magnitude stronger than those in Ref.~\cite{Raffelt}. We have also established new strong limits on the scalar-electron coupling $g_{\phi e}$ for the mass range $1\,\text{keV} < m_\phi < 210\,\text{MeV}$. Furthermore, we have improved the limits on the scalar-W-boson coupling by 2--5 orders of magnitude compared to Ref.~\cite{Stadnik2015}.

For light quarks ($u$, $d$, $s$), we have derived new limits for $m_\phi > 0.05\,\text{eV}$, while for heavy ones ($c$, $b$, $t$), we have established new limits across the entire mass range considered, $10^{-23}\,\text{eV} < m_\phi < 15\,\text{MeV}$. Additionally, our limits on the scalar-Higgs field interaction extend the results of Ref.~\cite{Stadnik2015} to scalar field masses $m_\phi > 10^{-14}\,\text{eV}$. 

These results represent significant advancements in constraining the interaction strengths of the scalar DM particle with Standard Model particles, providing tighter bounds across a broad range of scalar field masses.

It would be interesting to extend the results of this work to heavy scalar particles with masses $m>15$\,MeV. In this regime, particle accelerator and beam dump experiments are expected to provide strong constraints on new physics. However, such processes are primarily dominated by direct scalar particle production and annihilation, with loop quantum corrections playing a subdominant role. We also note that different particle accelerator and beam dump experiments have varying sensitivities to the scalar coupling with leptons and hadrons. Investigating this issue represents a separate study which is a subject of experimental work. Here, we focus on one-loop quantum corrections to the scalar-photon coupling for light scalar fields.

\vspace{2mm}
\textit{Acknowledgements.}--- 
We are grateful to Hooman Davoudiasl for stimulating discussions and to Edoardo Vitagliano for useful references. The work was supported by the Australian Research Council Grants No.\ DP230101058 and No.\ DP200100150.


\appendix
\section{Radiative corrections from charged meson loops}
\label{AppA}

Below the QCD scale, the QCD reduces to the effective theory of mesons. In this regime, the decay of the scalar field $\phi$ occurs via loops of charged $\pi$ and $K$ mesons in Fig.~\ref{figMesonLoops}. These mesons are pseudo Nambu-Goldstone bosons parametrizing the $[SU(3)_L\times SU(3)_R]/SU(3)_V$ coset space. This space may be described by the following unitary matrix
\begin{equation}
    U = \exp[i\sqrt{2}\Pi/f]\,,
    \label{U}
\end{equation}
where $f\approx 92$\,MeV is the pion decay constant and 
\begin{equation}
    \Pi = \left(
    \begin{array}{ccc}
     \frac{\pi^0}{\sqrt2} + \frac{\eta^8}{\sqrt{6}} & \pi^+ & K^+ \\
     \pi^- & -\frac{\pi^0}{\sqrt2} + \frac{\eta^8}{\sqrt6} & K^0 \\
     K^-   & \bar K^0 & \frac{-2\eta^8}{\sqrt6}
    \end{array}
    \right).
\end{equation}
The matrix (\ref{U}) transforms linearly under the $SU(3)_L\times SU(3)_R$ transformations, $U\to V_L U V_R$, where $V_{L(R)}\in SU(3)_{L(R)}$.

The QCD Lagrangian contains the following mass terms for the light $q=(u,d,s)$ quarks:
\begin{equation}
    {\cal L}_\mathrm{QCD} \supset - \bar q m q\,,
\end{equation}
where $m$ is the mass matrix $m=\mathrm{diag}(m_u,m_d,m_s)$. In the effective low-energy theory, quark masses give rise to the following mass terms of the charged pions:
\begin{equation}
    {\cal L}_\mathrm{eff} \supset m_\pi^2 \pi^+ \pi^- + m_K^2 K^+ K^-\,.
    \label{Lpions}
\end{equation}
Neglecting the QED corrections, the pion and kaon masses may be expressed via the quark masses:
\begin{equation}
\label{A5}
\begin{aligned}
    m_\pi^2 &\simeq B_0 (m_u + m_d)\,,\\
    m_K^2 &\simeq B_0(m_u + m_s)\,,
\end{aligned}
\end{equation}
where $B_0=O(1)$\,GeV is a constant which takes into account quark confinement.

The interaction of quarks with the background scalar field $\phi$ may be considered as the varying mass term in the QCD Lagrangian,
\begin{equation}
    {\cal L}_\mathrm{QCD}+{\cal L}_\mathrm{int}\supset 
    -\bar q M q\,,
    \label{LL}
\end{equation}
where $M = \mathrm{diag} (m_u + g_{\phi u}\phi,m_d + g_{\phi d}\phi,m_s + g_{\phi s}\phi)$. Similar to Eq.~(\ref{Lpions}), the low-energy theory of (\ref{LL}) may be described by the following effective Lagrangian
\begin{equation}
    \tilde{\cal L}_\mathrm{eff} = M_\pi^2 \pi^+ \pi^- 
    + M_K^2 K^+ K^-\,,
    \label{tildeLeff}
\end{equation}
where 
\begin{equation}
\begin{aligned}
    M_\pi^2 &= m_\pi^2 + B_0 (g_u + g_d)\phi\,,\\
    M_K^2 &= m_\pi^2 + B_0(g_u + g_s)\phi\,.
\end{aligned} 
\end{equation}
Thus, the interaction of the charged pions and kaons with the scalar field in Eq.~(\ref{tildeLeff}) may be cast in the form
\begin{equation}
\begin{aligned}
    \tilde {\cal L}_\mathrm{eff} &= m_\pi^2 \pi^+ \pi^- 
    + m_K^2 K^+ K^- \\
    &+g_{\phi\pi} \phi \pi^+ \pi^- 
    + g_{\phi K} \phi K^+ K^- \,,
\end{aligned}
\end{equation}
with scalar-meson interaction constants
\begin{equation}
\begin{aligned}
    g_{\phi \pi} &= B_0 (g_{\phi u} + g_{\phi d})\,,\\
    g_{\phi K} &= B_0 (g_{\phi u} + g_{\phi s})\,.
\end{aligned}
\label{A12}
\end{equation}

The contribution of one-loop diagrams in Fig.~\ref{figMesonLoops} to the scalar-photon coupling constant $g_{\phi\gamma}$ may be represented similar to Eq.~(\ref{result}):
\begin{equation}
    g_{\phi\gamma} = \frac{\alpha}{2\pi} \left(
    \frac{g_{\phi\pi}A_\pi}{m_\pi^2}
    +\frac{g_{\phi K}A_K}{m_K^2}
    \right)\,,
    \label{A14}
\end{equation}
with some dimensionless functions of the masses $A_{i}=A_{i}(m_\phi,m_{i})$, $i=\pi,K$. These functions were calculated in Ref.~\cite{Leutwyler89}:
\begin{equation}
    A_i =\frac1{2\tau_i} \left(
    \frac{\arcsin^2 \sqrt{\tau_i}}{\tau_i}-1
    \right)\,,
    \quad
    \tau_i\equiv\frac{m_\phi^2}{4m_i^2}\,.
    \label{Ai}
\end{equation}

In the limit $\tau_i\ll1$, the functions (\ref{Ai}) may be replaced by a constant,
\begin{equation}
    A_i|_{\tau\to0} \to \frac16\,.
\end{equation}
Thus, in the regime of the light scalar field the charged mesons contribute to the scalar-photon coupling (\ref{A14}) as
\begin{equation}
\begin{aligned}
    g_{\phi\gamma} &= \frac\alpha{12\pi}\left(
    \frac{g_{\phi\pi}}{m_{\pi}^2}
    +\frac{g_{\phi K}}{m_K^2}
    \right) \\
    &=\frac{\alpha B_0}{12\pi}\left[
    g_{\phi u} \left(\frac{1}{m_{\pi}^2}+\frac1{m_K^2}\right)
    +\frac{g_{\phi d}}{m_\pi^2}
    +\frac{g_{\phi s}}{m_K^2}
    \right],
\end{aligned}
\label{A16}
\end{equation}
where in the second line we applied the identities (\ref{A12}). Making use of Eqs.~(\ref{A5}), we eliminate the constant $B_0$ and the meson masses from Eq.~(\ref{A16})
\begin{equation}
    g_{\phi\gamma} = \frac{\alpha}{12\pi}
    \left[
    \frac{g_{\phi u}+g_{\phi s}+g_{\phi d}}{m_u+m_d}
    +\frac{g_{\phi u}+g_{\phi s}}{m_u + m_s}
    \right].
\end{equation}
Finally, substituting the values of the quark masses $m_u = 2.16$\,MeV, $m_d=4.67$\,MeV, and $m_s=93.4$\,MeV from Ref.~\cite{PDG}, we arrive at the relation (\ref{16}).

%
    

\begin{thebibliography}{81}%
    \makeatletter
    \providecommand \@ifxundefined [1]{%
     \@ifx{#1\undefined}
    }%
    \providecommand \@ifnum [1]{%
     \ifnum #1\expandafter \@firstoftwo
     \else \expandafter \@secondoftwo
     \fi
    }%
    \providecommand \@ifx [1]{%
     \ifx #1\expandafter \@firstoftwo
     \else \expandafter \@secondoftwo
     \fi
    }%
    \providecommand \natexlab [1]{#1}%
    \providecommand \enquote  [1]{``#1''}%
    \providecommand \bibnamefont  [1]{#1}%
    \providecommand \bibfnamefont [1]{#1}%
    \providecommand \citenamefont [1]{#1}%
    \providecommand \href@noop [0]{\@secondoftwo}%
    \providecommand \href [0]{\begingroup \@sanitize@url \@href}%
    \providecommand \@href[1]{\@@startlink{#1}\@@href}%
    \providecommand \@@href[1]{\endgroup#1\@@endlink}%
    \providecommand \@sanitize@url [0]{\catcode `\\12\catcode `\$12\catcode
      `\&12\catcode `\#12\catcode `\^12\catcode `\_12\catcode `\%12\relax}%
    \providecommand \@@startlink[1]{}%
    \providecommand \@@endlink[0]{}%
    \providecommand \url  [0]{\begingroup\@sanitize@url \@url }%
    \providecommand \@url [1]{\endgroup\@href {#1}{\urlprefix }}%
    \providecommand \urlprefix  [0]{URL }%
    \providecommand \Eprint [0]{\href }%
    \providecommand \doibase [0]{https://doi.org/}%
    \providecommand \selectlanguage [0]{\@gobble}%
    \providecommand \bibinfo  [0]{\@secondoftwo}%
    \providecommand \bibfield  [0]{\@secondoftwo}%
    \providecommand \translation [1]{[#1]}%
    \providecommand \BibitemOpen [0]{}%
    \providecommand \bibitemStop [0]{}%
    \providecommand \bibitemNoStop [0]{.\EOS\space}%
    \providecommand \EOS [0]{\spacefactor3000\relax}%
    \providecommand \BibitemShut  [1]{\csname bibitem#1\endcsname}%
    \let\auto@bib@innerbib\@empty
    \bibitem [{\citenamefont {Adams}\ \emph {et~al.}(2022)\citenamefont {Adams}
      \emph {et~al.}}]{SnowmassAxion}%
      \BibitemOpen
      \bibfield  {author} {\bibinfo {author} {\bibfnamefont {C.~B.}\ \bibnamefont
      {Adams}} \emph {et~al.},\ }\href@noop {} {\bibinfo {title} {{Axion Dark
      Matter}}} (\bibinfo {year} {2022}),\ \Eprint
      {https://arxiv.org/abs/2203.14923} {arXiv:2203.14923 [hep-ex]} \BibitemShut
      {NoStop}%
    \bibitem [{\citenamefont {Allison}\ \emph {et~al.}(2015)\citenamefont
      {Allison}, \citenamefont {Hill},\ and\ \citenamefont {Ross}}]{dilaton}%
      \BibitemOpen
      \bibfield  {author} {\bibinfo {author} {\bibfnamefont {K.}~\bibnamefont
      {Allison}}, \bibinfo {author} {\bibfnamefont {C.~T.}\ \bibnamefont {Hill}},\
      and\ \bibinfo {author} {\bibfnamefont {G.~G.}\ \bibnamefont {Ross}},\ }\href
      {https://doi.org/https://doi.org/10.1016/j.nuclphysb.2014.12.022} {\bibfield
      {journal} {\bibinfo  {journal} {Nucl. Phys. B}\ }\textbf {\bibinfo {volume}
      {891}},\ \bibinfo {pages} {613} (\bibinfo {year} {2015})},\ \Eprint
      {https://arxiv.org/abs/1409.4029} {arXiv:1409.4029 [hep-ph]} \BibitemShut
      {NoStop}%
    \bibitem [{\citenamefont {Taylor}\ and\ \citenamefont
      {Veneziano}(1988)}]{dilaton1}%
      \BibitemOpen
      \bibfield  {author} {\bibinfo {author} {\bibfnamefont {T.}~\bibnamefont
      {Taylor}}\ and\ \bibinfo {author} {\bibfnamefont {G.}~\bibnamefont
      {Veneziano}},\ }\href
      {https://doi.org/https://doi.org/10.1016/0370-2693(88)91290-7} {\bibfield
      {journal} {\bibinfo  {journal} {Phys. Lett. B}\ }\textbf {\bibinfo {volume}
      {213}},\ \bibinfo {pages} {450} (\bibinfo {year} {1988})}\BibitemShut
      {NoStop}%
    \bibitem [{\citenamefont {Damour}\ and\ \citenamefont
      {Polyakov}(1994{\natexlab{a}})}]{dilaton2}%
      \BibitemOpen
      \bibfield  {author} {\bibinfo {author} {\bibfnamefont {T.}~\bibnamefont
      {Damour}}\ and\ \bibinfo {author} {\bibfnamefont {A.}~\bibnamefont
      {Polyakov}},\ }\href
      {https://doi.org/https://doi.org/10.1016/0550-3213(94)90143-0} {\bibfield
      {journal} {\bibinfo  {journal} {Nucl. Phys. B}\ }\textbf {\bibinfo {volume}
      {423}},\ \bibinfo {pages} {532} (\bibinfo {year} {1994}{\natexlab{a}})},\
      \Eprint {https://arxiv.org/abs/hep-th/9401069} {arXiv:hep-th/9401069}
      \BibitemShut {NoStop}%
    \bibitem [{\citenamefont {Damour}\ and\ \citenamefont
      {Polyakov}(1994{\natexlab{b}})}]{dilaton3}%
      \BibitemOpen
      \bibfield  {author} {\bibinfo {author} {\bibfnamefont {T.}~\bibnamefont
      {Damour}}\ and\ \bibinfo {author} {\bibfnamefont {A.~M.}\ \bibnamefont
      {Polyakov}},\ }\href {https://doi.org/10.1007/BF02106709} {\bibfield
      {journal} {\bibinfo  {journal} {Gen. Rel. Grav.}\ }\textbf {\bibinfo {volume}
      {26}},\ \bibinfo {pages} {1171} (\bibinfo {year} {1994}{\natexlab{b}})},\
      \Eprint {https://arxiv.org/abs/gr-qc/9411069} {arXiv:gr-qc/9411069}
      \BibitemShut {NoStop}%
    \bibitem [{\citenamefont {Damour}\ \emph
      {et~al.}(2002{\natexlab{a}})\citenamefont {Damour}, \citenamefont {Piazza},\
      and\ \citenamefont {Veneziano}}]{dilaton4}%
      \BibitemOpen
      \bibfield  {author} {\bibinfo {author} {\bibfnamefont {T.}~\bibnamefont
      {Damour}}, \bibinfo {author} {\bibfnamefont {F.}~\bibnamefont {Piazza}},\
      and\ \bibinfo {author} {\bibfnamefont {G.}~\bibnamefont {Veneziano}},\ }\href
      {https://doi.org/10.1103/PhysRevLett.89.081601} {\bibfield  {journal}
      {\bibinfo  {journal} {Phys. Rev. Lett.}\ }\textbf {\bibinfo {volume} {89}},\
      \bibinfo {pages} {081601} (\bibinfo {year} {2002}{\natexlab{a}})},\ \Eprint
      {https://arxiv.org/abs/gr-qc/0204094} {arXiv:gr-qc/0204094} \BibitemShut
      {NoStop}%
    \bibitem [{\citenamefont {Damour}\ \emph
      {et~al.}(2002{\natexlab{b}})\citenamefont {Damour}, \citenamefont {Piazza},\
      and\ \citenamefont {Veneziano}}]{dilaton5}%
      \BibitemOpen
      \bibfield  {author} {\bibinfo {author} {\bibfnamefont {T.}~\bibnamefont
      {Damour}}, \bibinfo {author} {\bibfnamefont {F.}~\bibnamefont {Piazza}},\
      and\ \bibinfo {author} {\bibfnamefont {G.}~\bibnamefont {Veneziano}},\ }\href
      {https://doi.org/10.1103/PhysRevD.66.046007} {\bibfield  {journal} {\bibinfo
      {journal} {Phys. Rev. D}\ }\textbf {\bibinfo {volume} {66}},\ \bibinfo
      {pages} {046007} (\bibinfo {year} {2002}{\natexlab{b}})},\ \Eprint
      {https://arxiv.org/abs/hep-th/0205111} {arXiv:hep-th/0205111} \BibitemShut
      {NoStop}%
    \bibitem [{\citenamefont {Kaplan}\ and\ \citenamefont {Wise}(2000)}]{dilaton6}%
      \BibitemOpen
      \bibfield  {author} {\bibinfo {author} {\bibfnamefont {D.~B.}\ \bibnamefont
      {Kaplan}}\ and\ \bibinfo {author} {\bibfnamefont {M.~B.}\ \bibnamefont
      {Wise}},\ }\href {https://doi.org/10.1088/1126-6708/2000/08/037} {\bibfield
      {journal} {\bibinfo  {journal} {J. High Ener. Phys.}\ }\textbf {\bibinfo
      {volume} {2000}},\ \bibinfo {pages} {037} (\bibinfo {year} {2000})},\ \Eprint
      {https://arxiv.org/abs/hep-ph/0008116} {arXiv:hep-ph/0008116} \BibitemShut
      {NoStop}%
    \bibitem [{\citenamefont {Burrage}\ and\ \citenamefont
      {Sakstein}(2018)}]{chameleon}%
      \BibitemOpen
      \bibfield  {author} {\bibinfo {author} {\bibfnamefont {C.}~\bibnamefont
      {Burrage}}\ and\ \bibinfo {author} {\bibfnamefont {J.}~\bibnamefont
      {Sakstein}},\ }\href {https://doi.org/10.1007/s41114-018-0011-x} {\bibfield
      {journal} {\bibinfo  {journal} {Living Rev. Rel.}\ }\textbf {\bibinfo
      {volume} {21}},\ \bibinfo {pages} {1} (\bibinfo {year} {2018})},\ \Eprint
      {https://arxiv.org/abs/1709.09071} {arXiv:1709.09071 [astro-ph.CO]}
      \BibitemShut {NoStop}%
    \bibitem [{\citenamefont {Derevianko}\ and\ \citenamefont
      {Pospelov}(2014)}]{2phi1}%
      \BibitemOpen
      \bibfield  {author} {\bibinfo {author} {\bibfnamefont {A.}~\bibnamefont
      {Derevianko}}\ and\ \bibinfo {author} {\bibfnamefont {M.}~\bibnamefont
      {Pospelov}},\ }\href {https://doi.org/10.1038/nphys3137} {\bibfield
      {journal} {\bibinfo  {journal} {Nature Phys.}\ }\textbf {\bibinfo {volume}
      {10}},\ \bibinfo {pages} {933} (\bibinfo {year} {2014})},\ \Eprint
      {https://arxiv.org/abs/1311.1244} {arXiv:1311.1244 [physics.atom-ph]}
      \BibitemShut {NoStop}%
    \bibitem [{\citenamefont {Stadnik}\ and\ \citenamefont
      {Flambaum}(2015{\natexlab{a}})}]{Stadnik2015}%
      \BibitemOpen
      \bibfield  {author} {\bibinfo {author} {\bibfnamefont {Y.~V.}\ \bibnamefont
      {Stadnik}}\ and\ \bibinfo {author} {\bibfnamefont {V.~V.}\ \bibnamefont
      {Flambaum}},\ }\href {https://doi.org/10.1103/PhysRevLett.115.201301}
      {\bibfield  {journal} {\bibinfo  {journal} {Phys. Rev. Lett.}\ }\textbf
      {\bibinfo {volume} {115}},\ \bibinfo {pages} {201301} (\bibinfo {year}
      {2015}{\natexlab{a}})},\ \Eprint {https://arxiv.org/abs/1503.08540}
      {arXiv:1503.08540 [astro-ph.CO]} \BibitemShut {NoStop}%
    \bibitem [{\citenamefont {Stadnik}\ and\ \citenamefont
      {Flambaum}(2015{\natexlab{b}})}]{Stadnik}%
      \BibitemOpen
      \bibfield  {author} {\bibinfo {author} {\bibfnamefont {Y.~V.}\ \bibnamefont
      {Stadnik}}\ and\ \bibinfo {author} {\bibfnamefont {V.~V.}\ \bibnamefont
      {Flambaum}},\ }\href {https://doi.org/10.1103/PhysRevLett.114.161301}
      {\bibfield  {journal} {\bibinfo  {journal} {Phys. Rev. Lett.}\ }\textbf
      {\bibinfo {volume} {114}},\ \bibinfo {pages} {161301} (\bibinfo {year}
      {2015}{\natexlab{b}})},\ \Eprint {https://arxiv.org/abs/1412.7801}
      {arXiv:1412.7801 [hep-ph]} \BibitemShut {NoStop}%
    \bibitem [{\citenamefont {Antypas}\ \emph {et~al.}(2022)\citenamefont {Antypas}
      \emph {et~al.}}]{SnowmassScalar}%
      \BibitemOpen
      \bibfield  {author} {\bibinfo {author} {\bibfnamefont {D.}~\bibnamefont
      {Antypas}} \emph {et~al.},\ }\href@noop {} {\bibinfo {title} {{New Horizons:
      Scalar and Vector Ultralight Dark Matter}}} (\bibinfo {year} {2022}),\
      \Eprint {https://arxiv.org/abs/2203.14915} {arXiv:2203.14915 [hep-ex]}
      \BibitemShut {NoStop}%
    \bibitem [{\citenamefont {O'Hare}(2020)}]{AxionLimits}%
      \BibitemOpen
      \bibfield  {author} {\bibinfo {author} {\bibfnamefont {C.}~\bibnamefont
      {O'Hare}},\ }\href {https://doi.org/10.5281/zenodo.3932430} {\bibinfo {title}
      {cajohare/axionlimits: Axionlimits}},\ \bibinfo {howpublished}
      {\url{https://cajohare.github.io/AxionLimits/}} (\bibinfo {year}
      {2020})\BibitemShut {NoStop}%
    \bibitem [{\citenamefont {Ellis}\ \emph {et~al.}(1976)\citenamefont {Ellis},
      \citenamefont {Gaillard},\ and\ \citenamefont {Nanopoulos}}]{Ellis1975}%
      \BibitemOpen
      \bibfield  {author} {\bibinfo {author} {\bibfnamefont {J.~R.}\ \bibnamefont
      {Ellis}}, \bibinfo {author} {\bibfnamefont {M.~K.}\ \bibnamefont
      {Gaillard}},\ and\ \bibinfo {author} {\bibfnamefont {D.~V.}\ \bibnamefont
      {Nanopoulos}},\ }\href {https://doi.org/10.1016/0550-3213(76)90382-5}
      {\bibfield  {journal} {\bibinfo  {journal} {Nucl. Phys. B}\ }\textbf
      {\bibinfo {volume} {106}},\ \bibinfo {pages} {292} (\bibinfo {year}
      {1976})}\BibitemShut {NoStop}%
    \bibitem [{\citenamefont {Shifman}\ \emph {et~al.}(1979)\citenamefont
      {Shifman}, \citenamefont {Vainshtein}, \citenamefont {Voloshin},\ and\
      \citenamefont {Zakharov}}]{Shifman1979}%
      \BibitemOpen
      \bibfield  {author} {\bibinfo {author} {\bibfnamefont {M.~A.}\ \bibnamefont
      {Shifman}}, \bibinfo {author} {\bibfnamefont {A.~I.}\ \bibnamefont
      {Vainshtein}}, \bibinfo {author} {\bibfnamefont {M.~B.}\ \bibnamefont
      {Voloshin}},\ and\ \bibinfo {author} {\bibfnamefont {V.~I.}\ \bibnamefont
      {Zakharov}},\ }\href@noop {} {\bibfield  {journal} {\bibinfo  {journal} {Sov.
      J. Nucl. Phys.}\ }\textbf {\bibinfo {volume} {30}},\ \bibinfo {pages} {711}
      (\bibinfo {year} {1979})}\BibitemShut {NoStop}%
    \bibitem [{\citenamefont {Damour}\ and\ \citenamefont
      {Donoghue}(2010)}]{Damour}%
      \BibitemOpen
      \bibfield  {author} {\bibinfo {author} {\bibfnamefont {T.}~\bibnamefont
      {Damour}}\ and\ \bibinfo {author} {\bibfnamefont {J.~F.}\ \bibnamefont
      {Donoghue}},\ }\href {https://doi.org/10.1103/PhysRevD.82.084033} {\bibfield
      {journal} {\bibinfo  {journal} {Phys. Rev. D}\ }\textbf {\bibinfo {volume}
      {82}},\ \bibinfo {pages} {084033} (\bibinfo {year} {2010})},\ \Eprint
      {https://arxiv.org/abs/1007.2792} {arXiv:1007.2792 [gr-qc]} \BibitemShut
      {NoStop}%
    \bibitem [{\citenamefont {Arvanitaki}\ \emph {et~al.}(2015)\citenamefont
      {Arvanitaki}, \citenamefont {Huang},\ and\ \citenamefont
      {Van~Tilburg}}]{Arvanitaki}%
      \BibitemOpen
      \bibfield  {author} {\bibinfo {author} {\bibfnamefont {A.}~\bibnamefont
      {Arvanitaki}}, \bibinfo {author} {\bibfnamefont {J.}~\bibnamefont {Huang}},\
      and\ \bibinfo {author} {\bibfnamefont {K.}~\bibnamefont {Van~Tilburg}},\
      }\href {https://doi.org/10.1103/PhysRevD.91.015015} {\bibfield  {journal}
      {\bibinfo  {journal} {Phys. Rev. D}\ }\textbf {\bibinfo {volume} {91}},\
      \bibinfo {pages} {015015} (\bibinfo {year} {2015})},\ \Eprint
      {https://arxiv.org/abs/1405.2925} {arXiv:1405.2925 [hep-ph]} \BibitemShut
      {NoStop}%
    \bibitem [{\citenamefont {Djouadi}(2008)}]{HiggsReview}%
      \BibitemOpen
      \bibfield  {author} {\bibinfo {author} {\bibfnamefont {A.}~\bibnamefont
      {Djouadi}},\ }\href
      {https://doi.org/https://doi.org/10.1016/j.physrep.2007.10.004} {\bibfield
      {journal} {\bibinfo  {journal} {Phys. Rep.}\ }\textbf {\bibinfo {volume}
      {457}},\ \bibinfo {pages} {1} (\bibinfo {year} {2008})},\ \Eprint
      {https://arxiv.org/abs/hep-ph/0503172} {arXiv:hep-ph/0503172} \BibitemShut
      {NoStop}%
    \bibitem [{\citenamefont {Hees}\ \emph {et~al.}(2016)\citenamefont {Hees},
      \citenamefont {Gu\'ena}, \citenamefont {Abgrall}, \citenamefont {Bize},\ and\
      \citenamefont {Wolf}}]{phi-gamma1}%
      \BibitemOpen
      \bibfield  {author} {\bibinfo {author} {\bibfnamefont {A.}~\bibnamefont
      {Hees}}, \bibinfo {author} {\bibfnamefont {J.}~\bibnamefont {Gu\'ena}},
      \bibinfo {author} {\bibfnamefont {M.}~\bibnamefont {Abgrall}}, \bibinfo
      {author} {\bibfnamefont {S.}~\bibnamefont {Bize}},\ and\ \bibinfo {author}
      {\bibfnamefont {P.}~\bibnamefont {Wolf}},\ }\href
      {https://doi.org/10.1103/PhysRevLett.117.061301} {\bibfield  {journal}
      {\bibinfo  {journal} {Phys. Rev. Lett.}\ }\textbf {\bibinfo {volume} {117}},\
      \bibinfo {pages} {061301} (\bibinfo {year} {2016})},\ \Eprint
      {https://arxiv.org/abs/1604.08514} {arXiv:1604.08514 [gr-qc]} \BibitemShut
      {NoStop}%
    \bibitem [{\citenamefont {Filzinger}\ \emph
      {et~al.}(2023{\natexlab{a}})\citenamefont {Filzinger}, \citenamefont
      {D\"orscher}, \citenamefont {Lange}, \citenamefont {Klose}, \citenamefont
      {Steinel}, \citenamefont {Benkler}, \citenamefont {Peik}, \citenamefont
      {Lisdat},\ and\ \citenamefont {Huntemann}}]{phi-gamma2}%
      \BibitemOpen
      \bibfield  {author} {\bibinfo {author} {\bibfnamefont {M.}~\bibnamefont
      {Filzinger}}, \bibinfo {author} {\bibfnamefont {S.}~\bibnamefont
      {D\"orscher}}, \bibinfo {author} {\bibfnamefont {R.}~\bibnamefont {Lange}},
      \bibinfo {author} {\bibfnamefont {J.}~\bibnamefont {Klose}}, \bibinfo
      {author} {\bibfnamefont {M.}~\bibnamefont {Steinel}}, \bibinfo {author}
      {\bibfnamefont {E.}~\bibnamefont {Benkler}}, \bibinfo {author} {\bibfnamefont
      {E.}~\bibnamefont {Peik}}, \bibinfo {author} {\bibfnamefont {C.}~\bibnamefont
      {Lisdat}},\ and\ \bibinfo {author} {\bibfnamefont {N.}~\bibnamefont
      {Huntemann}},\ }\href {https://doi.org/10.1103/PhysRevLett.130.253001}
      {\bibfield  {journal} {\bibinfo  {journal} {Phys. Rev. Lett.}\ }\textbf
      {\bibinfo {volume} {130}},\ \bibinfo {pages} {253001} (\bibinfo {year}
      {2023}{\natexlab{a}})},\ \Eprint {https://arxiv.org/abs/2301.03433}
      {arXiv:2301.03433 [physics.atom-ph]} \BibitemShut {NoStop}%
    \bibitem [{\citenamefont {Sherrill}\ \emph {et~al.}(2023)\citenamefont
      {Sherrill} \emph {et~al.}}]{phi-gamma3}%
      \BibitemOpen
      \bibfield  {author} {\bibinfo {author} {\bibfnamefont {N.}~\bibnamefont
      {Sherrill}} \emph {et~al.},\ }\href
      {https://doi.org/10.1088/1367-2630/aceff6} {\bibfield  {journal} {\bibinfo
      {journal} {New J. Phys.}\ }\textbf {\bibinfo {volume} {25}},\ \bibinfo
      {pages} {093012} (\bibinfo {year} {2023})},\ \Eprint
      {https://arxiv.org/abs/2302.04565} {arXiv:2302.04565 [physics.atom-ph]}
      \BibitemShut {NoStop}%
    \bibitem [{\citenamefont {Berg\'e}\ \emph {et~al.}(2018)\citenamefont
      {Berg\'e}, \citenamefont {Brax}, \citenamefont {M\'etris}, \citenamefont
      {Pernot-Borr\`as}, \citenamefont {Touboul},\ and\ \citenamefont
      {Uzan}}]{phi-gamma4}%
      \BibitemOpen
      \bibfield  {author} {\bibinfo {author} {\bibfnamefont {J.}~\bibnamefont
      {Berg\'e}}, \bibinfo {author} {\bibfnamefont {P.}~\bibnamefont {Brax}},
      \bibinfo {author} {\bibfnamefont {G.}~\bibnamefont {M\'etris}}, \bibinfo
      {author} {\bibfnamefont {M.}~\bibnamefont {Pernot-Borr\`as}}, \bibinfo
      {author} {\bibfnamefont {P.}~\bibnamefont {Touboul}},\ and\ \bibinfo {author}
      {\bibfnamefont {J.-P.}\ \bibnamefont {Uzan}},\ }\href
      {https://doi.org/10.1103/PhysRevLett.120.141101} {\bibfield  {journal}
      {\bibinfo  {journal} {Phys. Rev. Lett.}\ }\textbf {\bibinfo {volume} {120}},\
      \bibinfo {pages} {141101} (\bibinfo {year} {2018})},\ \Eprint
      {https://arxiv.org/abs/1712.00483} {arXiv:1712.00483 [gr-qc]} \BibitemShut
      {NoStop}%
    \bibitem [{\citenamefont {Hees}\ \emph {et~al.}(2018)\citenamefont {Hees},
      \citenamefont {Minazzoli}, \citenamefont {Savalle}, \citenamefont {Stadnik},\
      and\ \citenamefont {Wolf}}]{phi-gamma5}%
      \BibitemOpen
      \bibfield  {author} {\bibinfo {author} {\bibfnamefont {A.}~\bibnamefont
      {Hees}}, \bibinfo {author} {\bibfnamefont {O.}~\bibnamefont {Minazzoli}},
      \bibinfo {author} {\bibfnamefont {E.}~\bibnamefont {Savalle}}, \bibinfo
      {author} {\bibfnamefont {Y.~V.}\ \bibnamefont {Stadnik}},\ and\ \bibinfo
      {author} {\bibfnamefont {P.}~\bibnamefont {Wolf}},\ }\href
      {https://doi.org/10.1103/PhysRevD.98.064051} {\bibfield  {journal} {\bibinfo
      {journal} {Phys. Rev. D}\ }\textbf {\bibinfo {volume} {98}},\ \bibinfo
      {pages} {064051} (\bibinfo {year} {2018})},\ \Eprint
      {https://arxiv.org/abs/1807.04512} {arXiv:1807.04512 [gr-qc]} \BibitemShut
      {NoStop}%
    \bibitem [{\citenamefont {Van~Tilburg}\ \emph {et~al.}(2015)\citenamefont
      {Van~Tilburg}, \citenamefont {Leefer}, \citenamefont {Bougas},\ and\
      \citenamefont {Budker}}]{phi-gamma6}%
      \BibitemOpen
      \bibfield  {author} {\bibinfo {author} {\bibfnamefont {K.}~\bibnamefont
      {Van~Tilburg}}, \bibinfo {author} {\bibfnamefont {N.}~\bibnamefont {Leefer}},
      \bibinfo {author} {\bibfnamefont {L.}~\bibnamefont {Bougas}},\ and\ \bibinfo
      {author} {\bibfnamefont {D.}~\bibnamefont {Budker}},\ }\href
      {https://doi.org/10.1103/PhysRevLett.115.011802} {\bibfield  {journal}
      {\bibinfo  {journal} {Phys. Rev. Lett.}\ }\textbf {\bibinfo {volume} {115}},\
      \bibinfo {pages} {011802} (\bibinfo {year} {2015})},\ \Eprint
      {https://arxiv.org/abs/1503.06886} {arXiv:1503.06886 [physics.atom-ph]}
      \BibitemShut {NoStop}%
    \bibitem [{\citenamefont {Beloy}\ \emph {et~al.}(2021)\citenamefont {Beloy}
      \emph {et~al.}}]{phi-gamma7}%
      \BibitemOpen
      \bibfield  {author} {\bibinfo {author} {\bibfnamefont {K.}~\bibnamefont
      {Beloy}} \emph {et~al.} (\bibinfo {collaboration} {BACON}),\ }\href
      {https://doi.org/10.1038/s41586-021-03253-4} {\bibfield  {journal} {\bibinfo
      {journal} {Nature}\ }\textbf {\bibinfo {volume} {591}},\ \bibinfo {pages}
      {564} (\bibinfo {year} {2021})},\ \Eprint {https://arxiv.org/abs/2005.14694}
      {arXiv:2005.14694 [physics.atom-ph]} \BibitemShut {NoStop}%
    \bibitem [{\citenamefont {Kennedy}\ \emph {et~al.}(2020)\citenamefont
      {Kennedy}, \citenamefont {Oelker}, \citenamefont {Robinson}, \citenamefont
      {Bothwell}, \citenamefont {Kedar}, \citenamefont {Milner}, \citenamefont
      {Marti}, \citenamefont {Derevianko},\ and\ \citenamefont {Ye}}]{phi-gamma8}%
      \BibitemOpen
      \bibfield  {author} {\bibinfo {author} {\bibfnamefont {C.~J.}\ \bibnamefont
      {Kennedy}}, \bibinfo {author} {\bibfnamefont {E.}~\bibnamefont {Oelker}},
      \bibinfo {author} {\bibfnamefont {J.~M.}\ \bibnamefont {Robinson}}, \bibinfo
      {author} {\bibfnamefont {T.}~\bibnamefont {Bothwell}}, \bibinfo {author}
      {\bibfnamefont {D.}~\bibnamefont {Kedar}}, \bibinfo {author} {\bibfnamefont
      {W.~R.}\ \bibnamefont {Milner}}, \bibinfo {author} {\bibfnamefont {G.~E.}\
      \bibnamefont {Marti}}, \bibinfo {author} {\bibfnamefont {A.}~\bibnamefont
      {Derevianko}},\ and\ \bibinfo {author} {\bibfnamefont {J.}~\bibnamefont
      {Ye}},\ }\href {https://doi.org/10.1103/PhysRevLett.125.201302} {\bibfield
      {journal} {\bibinfo  {journal} {Phys. Rev. Lett.}\ }\textbf {\bibinfo
      {volume} {125}},\ \bibinfo {pages} {201302} (\bibinfo {year} {2020})},\
      \Eprint {https://arxiv.org/abs/2008.08773} {arXiv:2008.08773
      [physics.atom-ph]} \BibitemShut {NoStop}%
    \bibitem [{\citenamefont {Adelberger}\ \emph {et~al.}(2003)\citenamefont
      {Adelberger}, \citenamefont {Heckel},\ and\ \citenamefont
      {Nelson}}]{phi-gamma9}%
      \BibitemOpen
      \bibfield  {author} {\bibinfo {author} {\bibfnamefont {E.~G.}\ \bibnamefont
      {Adelberger}}, \bibinfo {author} {\bibfnamefont {B.~R.}\ \bibnamefont
      {Heckel}},\ and\ \bibinfo {author} {\bibfnamefont {A.~E.}\ \bibnamefont
      {Nelson}},\ }\href {https://doi.org/10.1146/annurev.nucl.53.041002.110503}
      {\bibfield  {journal} {\bibinfo  {journal} {Ann. Rev. Nucl. Part. Sci.}\
      }\textbf {\bibinfo {volume} {53}},\ \bibinfo {pages} {77} (\bibinfo {year}
      {2003})},\ \Eprint {https://arxiv.org/abs/hep-ph/0307284}
      {arXiv:hep-ph/0307284} \BibitemShut {NoStop}%
    \bibitem [{\citenamefont {Fischbach}\ and\ \citenamefont
      {Talmadge}(1996)}]{phi-gamma10}%
      \BibitemOpen
      \bibfield  {author} {\bibinfo {author} {\bibfnamefont {E.}~\bibnamefont
      {Fischbach}}\ and\ \bibinfo {author} {\bibfnamefont {C.}~\bibnamefont
      {Talmadge}},\ }in\ \href@noop {} {\emph {\bibinfo {booktitle} {{31st
      Rencontres de Moriond: Dark Matter and Cosmology, Quantum Measurements and
      Experimental Gravitation}}}}\ (\bibinfo {year} {1996})\ pp.\ \bibinfo {pages}
      {443--451},\ \Eprint {https://arxiv.org/abs/hep-ph/9606249}
      {arXiv:hep-ph/9606249} \BibitemShut {NoStop}%
    \bibitem [{\citenamefont {Campbell}\ \emph {et~al.}(2021)\citenamefont
      {Campbell}, \citenamefont {McAllister}, \citenamefont {Goryachev},
      \citenamefont {Ivanov},\ and\ \citenamefont {Tobar}}]{phi-gamma11}%
      \BibitemOpen
      \bibfield  {author} {\bibinfo {author} {\bibfnamefont {W.~M.}\ \bibnamefont
      {Campbell}}, \bibinfo {author} {\bibfnamefont {B.~T.}\ \bibnamefont
      {McAllister}}, \bibinfo {author} {\bibfnamefont {M.}~\bibnamefont
      {Goryachev}}, \bibinfo {author} {\bibfnamefont {E.~N.}\ \bibnamefont
      {Ivanov}},\ and\ \bibinfo {author} {\bibfnamefont {M.~E.}\ \bibnamefont
      {Tobar}},\ }\href {https://doi.org/10.1103/PhysRevLett.126.071301} {\bibfield
       {journal} {\bibinfo  {journal} {Phys. Rev. Lett.}\ }\textbf {\bibinfo
      {volume} {126}},\ \bibinfo {pages} {071301} (\bibinfo {year} {2021})},\
      \Eprint {https://arxiv.org/abs/2010.08107} {arXiv:2010.08107 [hep-ex]}
      \BibitemShut {NoStop}%
    \bibitem [{\citenamefont {Zhang}\ \emph {et~al.}(2023)\citenamefont {Zhang},
      \citenamefont {Banerjee}, \citenamefont {Leyser}, \citenamefont {Perez},
      \citenamefont {Schiller}, \citenamefont {Budker},\ and\ \citenamefont
      {Antypas}}]{phi-gamma12}%
      \BibitemOpen
      \bibfield  {author} {\bibinfo {author} {\bibfnamefont {X.}~\bibnamefont
      {Zhang}}, \bibinfo {author} {\bibfnamefont {A.}~\bibnamefont {Banerjee}},
      \bibinfo {author} {\bibfnamefont {M.}~\bibnamefont {Leyser}}, \bibinfo
      {author} {\bibfnamefont {G.}~\bibnamefont {Perez}}, \bibinfo {author}
      {\bibfnamefont {S.}~\bibnamefont {Schiller}}, \bibinfo {author}
      {\bibfnamefont {D.}~\bibnamefont {Budker}},\ and\ \bibinfo {author}
      {\bibfnamefont {D.}~\bibnamefont {Antypas}},\ }\href
      {https://doi.org/10.1103/PhysRevLett.130.251002} {\bibfield  {journal}
      {\bibinfo  {journal} {Phys. Rev. Lett.}\ }\textbf {\bibinfo {volume} {130}},\
      \bibinfo {pages} {251002} (\bibinfo {year} {2023})},\ \Eprint
      {https://arxiv.org/abs/2212.04413} {arXiv:2212.04413 [physics.atom-ph]}
      \BibitemShut {NoStop}%
    \bibitem [{\citenamefont {Aharony}\ \emph {et~al.}(2021)\citenamefont
      {Aharony}, \citenamefont {Akerman}, \citenamefont {Ozeri}, \citenamefont
      {Perez}, \citenamefont {Savoray},\ and\ \citenamefont
      {Shaniv}}]{phi-gamma13}%
      \BibitemOpen
      \bibfield  {author} {\bibinfo {author} {\bibfnamefont {S.}~\bibnamefont
      {Aharony}}, \bibinfo {author} {\bibfnamefont {N.}~\bibnamefont {Akerman}},
      \bibinfo {author} {\bibfnamefont {R.}~\bibnamefont {Ozeri}}, \bibinfo
      {author} {\bibfnamefont {G.}~\bibnamefont {Perez}}, \bibinfo {author}
      {\bibfnamefont {I.}~\bibnamefont {Savoray}},\ and\ \bibinfo {author}
      {\bibfnamefont {R.}~\bibnamefont {Shaniv}},\ }\href
      {https://doi.org/10.1103/PhysRevD.103.075017} {\bibfield  {journal} {\bibinfo
       {journal} {Phys. Rev. D}\ }\textbf {\bibinfo {volume} {103}},\ \bibinfo
      {pages} {075017} (\bibinfo {year} {2021})},\ \Eprint
      {https://arxiv.org/abs/1902.02788} {arXiv:1902.02788 [hep-ph]} \BibitemShut
      {NoStop}%
    \bibitem [{\citenamefont {Oswald}\ \emph {et~al.}(2022)\citenamefont {Oswald}
      \emph {et~al.}}]{phi-gamma14}%
      \BibitemOpen
      \bibfield  {author} {\bibinfo {author} {\bibfnamefont {R.}~\bibnamefont
      {Oswald}} \emph {et~al.},\ }\href
      {https://doi.org/10.1103/PhysRevLett.129.031302} {\bibfield  {journal}
      {\bibinfo  {journal} {Phys. Rev. Lett.}\ }\textbf {\bibinfo {volume} {129}},\
      \bibinfo {pages} {031302} (\bibinfo {year} {2022})},\ \Eprint
      {https://arxiv.org/abs/2111.06883} {arXiv:2111.06883 [hep-ph]} \BibitemShut
      {NoStop}%
    \bibitem [{\citenamefont {Fukusumi}\ \emph {et~al.}(2023)\citenamefont
      {Fukusumi}, \citenamefont {Morisaki},\ and\ \citenamefont
      {Suyama}}]{phi-gamma15}%
      \BibitemOpen
      \bibfield  {author} {\bibinfo {author} {\bibfnamefont {K.}~\bibnamefont
      {Fukusumi}}, \bibinfo {author} {\bibfnamefont {S.}~\bibnamefont {Morisaki}},\
      and\ \bibinfo {author} {\bibfnamefont {T.}~\bibnamefont {Suyama}},\ }\href
      {https://doi.org/10.1103/PhysRevD.108.095054} {\bibfield  {journal} {\bibinfo
       {journal} {Phys. Rev. D}\ }\textbf {\bibinfo {volume} {108}},\ \bibinfo
      {pages} {095054} (\bibinfo {year} {2023})},\ \Eprint
      {https://arxiv.org/abs/2303.13088} {arXiv:2303.13088 [hep-ph]} \BibitemShut
      {NoStop}%
    \bibitem [{\citenamefont {Vermeulen}\ \emph {et~al.}(2021)\citenamefont
      {Vermeulen} \emph {et~al.}}]{phi-gamma16}%
      \BibitemOpen
      \bibfield  {author} {\bibinfo {author} {\bibfnamefont {S.}~\bibnamefont
      {Vermeulen}} \emph {et~al.},\ }\href
      {https://doi.org/10.1038/s41586-021-04031-y} {\bibfield  {journal} {\bibinfo
      {journal} {Nature}\ }\textbf {\bibinfo {volume} {600}},\ \bibinfo {pages}
      {424} (\bibinfo {year} {2021})},\ \Eprint {https://arxiv.org/abs/2103.03783}
      {arXiv:2103.03783 [gr-qc]} \BibitemShut {NoStop}%
    \bibitem [{\citenamefont {Savalle}\ \emph {et~al.}(2021)\citenamefont
      {Savalle}, \citenamefont {Hees}, \citenamefont {Frank}, \citenamefont
      {Cantin}, \citenamefont {Pottie}, \citenamefont {Roberts}, \citenamefont
      {Cros}, \citenamefont {Mcallister},\ and\ \citenamefont
      {Wolf}}]{phi-gamma17}%
      \BibitemOpen
      \bibfield  {author} {\bibinfo {author} {\bibfnamefont {E.}~\bibnamefont
      {Savalle}}, \bibinfo {author} {\bibfnamefont {A.}~\bibnamefont {Hees}},
      \bibinfo {author} {\bibfnamefont {F.}~\bibnamefont {Frank}}, \bibinfo
      {author} {\bibfnamefont {E.}~\bibnamefont {Cantin}}, \bibinfo {author}
      {\bibfnamefont {P.-E.}\ \bibnamefont {Pottie}}, \bibinfo {author}
      {\bibfnamefont {B.~M.}\ \bibnamefont {Roberts}}, \bibinfo {author}
      {\bibfnamefont {L.}~\bibnamefont {Cros}}, \bibinfo {author} {\bibfnamefont
      {B.~T.}\ \bibnamefont {Mcallister}},\ and\ \bibinfo {author} {\bibfnamefont
      {P.}~\bibnamefont {Wolf}},\ }\href
      {https://doi.org/10.1103/PhysRevLett.126.051301} {\bibfield  {journal}
      {\bibinfo  {journal} {Phys. Rev. Lett.}\ }\textbf {\bibinfo {volume} {126}},\
      \bibinfo {pages} {051301} (\bibinfo {year} {2021})},\ \Eprint
      {https://arxiv.org/abs/2006.07055} {arXiv:2006.07055 [gr-qc]} \BibitemShut
      {NoStop}%
    \bibitem [{\citenamefont {Aiello}\ \emph {et~al.}(2022)\citenamefont {Aiello},
      \citenamefont {Richardson}, \citenamefont {Vermeulen}, \citenamefont {Grote},
      \citenamefont {Hogan}, \citenamefont {Kwon},\ and\ \citenamefont
      {Stoughton}}]{phi-gamma18}%
      \BibitemOpen
      \bibfield  {author} {\bibinfo {author} {\bibfnamefont {L.}~\bibnamefont
      {Aiello}}, \bibinfo {author} {\bibfnamefont {J.~W.}\ \bibnamefont
      {Richardson}}, \bibinfo {author} {\bibfnamefont {S.~M.}\ \bibnamefont
      {Vermeulen}}, \bibinfo {author} {\bibfnamefont {H.}~\bibnamefont {Grote}},
      \bibinfo {author} {\bibfnamefont {C.}~\bibnamefont {Hogan}}, \bibinfo
      {author} {\bibfnamefont {O.}~\bibnamefont {Kwon}},\ and\ \bibinfo {author}
      {\bibfnamefont {C.}~\bibnamefont {Stoughton}},\ }\href
      {https://doi.org/10.1103/PhysRevLett.128.121101} {\bibfield  {journal}
      {\bibinfo  {journal} {Phys. Rev. Lett.}\ }\textbf {\bibinfo {volume} {128}},\
      \bibinfo {pages} {121101} (\bibinfo {year} {2022})},\ \Eprint
      {https://arxiv.org/abs/2108.04746} {arXiv:2108.04746 [gr-qc]} \BibitemShut
      {NoStop}%
    \bibitem [{\citenamefont {Tretiak}\ \emph {et~al.}(2022)\citenamefont
      {Tretiak}, \citenamefont {Zhang}, \citenamefont {Figueroa}, \citenamefont
      {Antypas}, \citenamefont {Brogna}, \citenamefont {Banerjee}, \citenamefont
      {Perez},\ and\ \citenamefont {Budker}}]{phi-gamma19}%
      \BibitemOpen
      \bibfield  {author} {\bibinfo {author} {\bibfnamefont {O.}~\bibnamefont
      {Tretiak}}, \bibinfo {author} {\bibfnamefont {X.}~\bibnamefont {Zhang}},
      \bibinfo {author} {\bibfnamefont {N.~L.}\ \bibnamefont {Figueroa}}, \bibinfo
      {author} {\bibfnamefont {D.}~\bibnamefont {Antypas}}, \bibinfo {author}
      {\bibfnamefont {A.}~\bibnamefont {Brogna}}, \bibinfo {author} {\bibfnamefont
      {A.}~\bibnamefont {Banerjee}}, \bibinfo {author} {\bibfnamefont
      {G.}~\bibnamefont {Perez}},\ and\ \bibinfo {author} {\bibfnamefont
      {D.}~\bibnamefont {Budker}},\ }\href
      {https://doi.org/10.1103/PhysRevLett.129.031301} {\bibfield  {journal}
      {\bibinfo  {journal} {Phys. Rev. Lett.}\ }\textbf {\bibinfo {volume} {129}},\
      \bibinfo {pages} {031301} (\bibinfo {year} {2022})},\ \Eprint
      {https://arxiv.org/abs/2201.02042} {arXiv:2201.02042 [hep-ph]} \BibitemShut
      {NoStop}%
    \bibitem [{\citenamefont {Flambaum}\ \emph {et~al.}(2023)\citenamefont
      {Flambaum}, \citenamefont {Mansour}, \citenamefont {Samsonov},\ and\
      \citenamefont {Weitenberg}}]{my1}%
      \BibitemOpen
      \bibfield  {author} {\bibinfo {author} {\bibfnamefont {V.~V.}\ \bibnamefont
      {Flambaum}}, \bibinfo {author} {\bibfnamefont {A.~J.}\ \bibnamefont
      {Mansour}}, \bibinfo {author} {\bibfnamefont {I.~B.}\ \bibnamefont
      {Samsonov}},\ and\ \bibinfo {author} {\bibfnamefont {C.}~\bibnamefont
      {Weitenberg}},\ }\href {https://doi.org/10.1103/PhysRevD.107.015008}
      {\bibfield  {journal} {\bibinfo  {journal} {Phys. Rev. D}\ }\textbf {\bibinfo
      {volume} {107}},\ \bibinfo {pages} {015008} (\bibinfo {year} {2023})},\
      \Eprint {https://arxiv.org/abs/2210.08778} {arXiv:2210.08778 [hep-ph]}
      \BibitemShut {NoStop}%
    \bibitem [{\citenamefont {Bloch}\ \emph {et~al.}(2023)\citenamefont {Bloch},
      \citenamefont {Budker}, \citenamefont {Flambaum}, \citenamefont {Samsonov},
      \citenamefont {Sushkov},\ and\ \citenamefont {Tretiak}}]{my2}%
      \BibitemOpen
      \bibfield  {author} {\bibinfo {author} {\bibfnamefont {I.~M.}\ \bibnamefont
      {Bloch}}, \bibinfo {author} {\bibfnamefont {D.}~\bibnamefont {Budker}},
      \bibinfo {author} {\bibfnamefont {V.~V.}\ \bibnamefont {Flambaum}}, \bibinfo
      {author} {\bibfnamefont {I.~B.}\ \bibnamefont {Samsonov}}, \bibinfo {author}
      {\bibfnamefont {A.~O.}\ \bibnamefont {Sushkov}},\ and\ \bibinfo {author}
      {\bibfnamefont {O.}~\bibnamefont {Tretiak}},\ }\href
      {https://doi.org/10.1103/PhysRevD.107.075033} {\bibfield  {journal} {\bibinfo
       {journal} {Phys. Rev. D}\ }\textbf {\bibinfo {volume} {107}},\ \bibinfo
      {pages} {075033} (\bibinfo {year} {2023})},\ \Eprint
      {https://arxiv.org/abs/2301.08514} {arXiv:2301.08514 [hep-ph]} \BibitemShut
      {NoStop}%
    \bibitem [{\citenamefont {Branca}\ \emph {et~al.}(2017)\citenamefont {Branca}
      \emph {et~al.}}]{AURIGA}%
      \BibitemOpen
      \bibfield  {author} {\bibinfo {author} {\bibfnamefont {A.}~\bibnamefont
      {Branca}} \emph {et~al.},\ }\href
      {https://doi.org/10.1103/PhysRevLett.118.021302} {\bibfield  {journal}
      {\bibinfo  {journal} {Phys. Rev. Lett.}\ }\textbf {\bibinfo {volume} {118}},\
      \bibinfo {pages} {021302} (\bibinfo {year} {2017})},\ \Eprint
      {https://arxiv.org/abs/1607.07327} {arXiv:1607.07327 [hep-ex]} \BibitemShut
      {NoStop}%
    \bibitem [{\citenamefont {Filzinger}\ \emph
      {et~al.}(2023{\natexlab{b}})\citenamefont {Filzinger}, \citenamefont
      {Caddell}, \citenamefont {Jani}, \citenamefont {Steinel}, \citenamefont
      {Giani}, \citenamefont {Huntemann},\ and\ \citenamefont
      {Roberts}}]{Cavities}%
      \BibitemOpen
      \bibfield  {author} {\bibinfo {author} {\bibfnamefont {M.}~\bibnamefont
      {Filzinger}}, \bibinfo {author} {\bibfnamefont {A.~R.}\ \bibnamefont
      {Caddell}}, \bibinfo {author} {\bibfnamefont {D.}~\bibnamefont {Jani}},
      \bibinfo {author} {\bibfnamefont {M.}~\bibnamefont {Steinel}}, \bibinfo
      {author} {\bibfnamefont {L.}~\bibnamefont {Giani}}, \bibinfo {author}
      {\bibfnamefont {N.}~\bibnamefont {Huntemann}},\ and\ \bibinfo {author}
      {\bibfnamefont {B.~M.}\ \bibnamefont {Roberts}},\ }\href@noop {} {\bibinfo
      {title} {{Ultralight dark matter search with space-time separated atomic
      clocks and cavities}}} (\bibinfo {year} {2023}{\natexlab{b}}),\ \Eprint
      {https://arxiv.org/abs/2312.13723} {arXiv:2312.13723 [hep-ph]} \BibitemShut
      {NoStop}%
    \bibitem [{\citenamefont {Kobayashi}\ \emph {et~al.}(2022)\citenamefont
      {Kobayashi} \emph {et~al.}}]{YbCs}%
      \BibitemOpen
      \bibfield  {author} {\bibinfo {author} {\bibfnamefont {T.}~\bibnamefont
      {Kobayashi}} \emph {et~al.},\ }\href
      {https://doi.org/10.1103/PhysRevLett.129.241301} {\bibfield  {journal}
      {\bibinfo  {journal} {Phys. Rev. Lett.}\ }\textbf {\bibinfo {volume} {129}},\
      \bibinfo {pages} {241301} (\bibinfo {year} {2022})},\ \Eprint
      {https://arxiv.org/abs/2212.05721} {arXiv:2212.05721 [physics.atom-ph]}
      \BibitemShut {NoStop}%
    \bibitem [{\citenamefont {Afzal}\ \emph {et~al.}(2023)\citenamefont {Afzal}
      \emph {et~al.}}]{NANOGrav}%
      \BibitemOpen
      \bibfield  {author} {\bibinfo {author} {\bibfnamefont {A.}~\bibnamefont
      {Afzal}} \emph {et~al.} (\bibinfo {collaboration} {NANOGrav}),\ }\href
      {https://doi.org/10.3847/2041-8213/acdc91} {\bibfield  {journal} {\bibinfo
      {journal} {Astrophys. J. Lett.}\ }\textbf {\bibinfo {volume} {951}},\
      \bibinfo {pages} {L11} (\bibinfo {year} {2023})},\ \Eprint
      {https://arxiv.org/abs/2306.16219} {arXiv:2306.16219 [astro-ph.HE]}
      \BibitemShut {NoStop}%
    \bibitem [{\citenamefont {Cadamuro}\ and\ \citenamefont
      {Redondo}(2012)}]{XRAY}%
      \BibitemOpen
      \bibfield  {author} {\bibinfo {author} {\bibfnamefont {D.}~\bibnamefont
      {Cadamuro}}\ and\ \bibinfo {author} {\bibfnamefont {J.}~\bibnamefont
      {Redondo}},\ }\href {https://doi.org/10.1088/1475-7516/2012/02/032}
      {\bibfield  {journal} {\bibinfo  {journal} {J. Cosm. Astropart. Phys.}\
      }\textbf {\bibinfo {volume} {2012}},\ \bibinfo {pages} {032} (\bibinfo {year}
      {2012})},\ \Eprint {https://arxiv.org/abs/1110.2895} {arXiv:1110.2895
      [hep-ph]} \BibitemShut {NoStop}%
    \bibitem [{\citenamefont {Foster}\ \emph {et~al.}(2021)\citenamefont {Foster},
      \citenamefont {Kongsore}, \citenamefont {Dessert}, \citenamefont {Park},
      \citenamefont {Rodd}, \citenamefont {Cranmer},\ and\ \citenamefont
      {Safdi}}]{XMM}%
      \BibitemOpen
      \bibfield  {author} {\bibinfo {author} {\bibfnamefont {J.~W.}\ \bibnamefont
      {Foster}}, \bibinfo {author} {\bibfnamefont {M.}~\bibnamefont {Kongsore}},
      \bibinfo {author} {\bibfnamefont {C.}~\bibnamefont {Dessert}}, \bibinfo
      {author} {\bibfnamefont {Y.}~\bibnamefont {Park}}, \bibinfo {author}
      {\bibfnamefont {N.~L.}\ \bibnamefont {Rodd}}, \bibinfo {author}
      {\bibfnamefont {K.}~\bibnamefont {Cranmer}},\ and\ \bibinfo {author}
      {\bibfnamefont {B.~R.}\ \bibnamefont {Safdi}},\ }\href
      {https://doi.org/10.1103/PhysRevLett.127.051101} {\bibfield  {journal}
      {\bibinfo  {journal} {Phys. Rev. Lett.}\ }\textbf {\bibinfo {volume} {127}},\
      \bibinfo {pages} {051101} (\bibinfo {year} {2021})},\ \Eprint
      {https://arxiv.org/abs/2102.02207} {arXiv:2102.02207 [astro-ph.CO]}
      \BibitemShut {NoStop}%
    \bibitem [{\citenamefont {Perez}\ \emph {et~al.}(2017)\citenamefont {Perez},
      \citenamefont {Ng}, \citenamefont {Beacom}, \citenamefont {Hersh},
      \citenamefont {Horiuchi},\ and\ \citenamefont {Krivonos}}]{NuStar1}%
      \BibitemOpen
      \bibfield  {author} {\bibinfo {author} {\bibfnamefont {K.}~\bibnamefont
      {Perez}}, \bibinfo {author} {\bibfnamefont {K.~C.~Y.}\ \bibnamefont {Ng}},
      \bibinfo {author} {\bibfnamefont {J.~F.}\ \bibnamefont {Beacom}}, \bibinfo
      {author} {\bibfnamefont {C.}~\bibnamefont {Hersh}}, \bibinfo {author}
      {\bibfnamefont {S.}~\bibnamefont {Horiuchi}},\ and\ \bibinfo {author}
      {\bibfnamefont {R.}~\bibnamefont {Krivonos}},\ }\href
      {https://doi.org/10.1103/PhysRevD.95.123002} {\bibfield  {journal} {\bibinfo
      {journal} {Phys. Rev. D}\ }\textbf {\bibinfo {volume} {95}},\ \bibinfo
      {pages} {123002} (\bibinfo {year} {2017})},\ \Eprint
      {https://arxiv.org/abs/1609.00667} {arXiv:1609.00667 [astro-ph.HE]}
      \BibitemShut {NoStop}%
    \bibitem [{\citenamefont {Roach}\ \emph {et~al.}(2023)\citenamefont {Roach}
      \emph {et~al.}}]{NuStar2}%
      \BibitemOpen
      \bibfield  {author} {\bibinfo {author} {\bibfnamefont {B.~M.}\ \bibnamefont
      {Roach}} \emph {et~al.},\ }\href
      {https://doi.org/10.1103/PhysRevD.107.023009} {\bibfield  {journal} {\bibinfo
       {journal} {Phys. Rev. D}\ }\textbf {\bibinfo {volume} {107}},\ \bibinfo
      {pages} {023009} (\bibinfo {year} {2023})},\ \Eprint
      {https://arxiv.org/abs/2207.04572} {arXiv:2207.04572 [astro-ph.HE]}
      \BibitemShut {NoStop}%
    \bibitem [{\citenamefont {Ng}\ \emph {et~al.}(2019)\citenamefont {Ng},
      \citenamefont {Roach}, \citenamefont {Perez}, \citenamefont {Beacom},
      \citenamefont {Horiuchi}, \citenamefont {Krivonos},\ and\ \citenamefont
      {Wik}}]{NuStar3}%
      \BibitemOpen
      \bibfield  {author} {\bibinfo {author} {\bibfnamefont {K.~C.~Y.}\
      \bibnamefont {Ng}}, \bibinfo {author} {\bibfnamefont {B.~M.}\ \bibnamefont
      {Roach}}, \bibinfo {author} {\bibfnamefont {K.}~\bibnamefont {Perez}},
      \bibinfo {author} {\bibfnamefont {J.~F.}\ \bibnamefont {Beacom}}, \bibinfo
      {author} {\bibfnamefont {S.}~\bibnamefont {Horiuchi}}, \bibinfo {author}
      {\bibfnamefont {R.}~\bibnamefont {Krivonos}},\ and\ \bibinfo {author}
      {\bibfnamefont {D.~R.}\ \bibnamefont {Wik}},\ }\href
      {https://doi.org/10.1103/PhysRevD.99.083005} {\bibfield  {journal} {\bibinfo
      {journal} {Phys. Rev. D}\ }\textbf {\bibinfo {volume} {99}},\ \bibinfo
      {pages} {083005} (\bibinfo {year} {2019})},\ \Eprint
      {https://arxiv.org/abs/1901.01262} {arXiv:1901.01262 [astro-ph.HE]}
      \BibitemShut {NoStop}%
    \bibitem [{\citenamefont {Calore}\ \emph {et~al.}(2023)\citenamefont {Calore},
      \citenamefont {Dekker}, \citenamefont {Serpico},\ and\ \citenamefont
      {Siegert}}]{INTEGRAL}%
      \BibitemOpen
      \bibfield  {author} {\bibinfo {author} {\bibfnamefont {F.}~\bibnamefont
      {Calore}}, \bibinfo {author} {\bibfnamefont {A.}~\bibnamefont {Dekker}},
      \bibinfo {author} {\bibfnamefont {P.~D.}\ \bibnamefont {Serpico}},\ and\
      \bibinfo {author} {\bibfnamefont {T.}~\bibnamefont {Siegert}},\ }\href
      {https://doi.org/10.1093/mnras/stad457} {\bibfield  {journal} {\bibinfo
      {journal} {Mon. Not. Roy. Astron. Soc.}\ }\textbf {\bibinfo {volume} {520}},\
      \bibinfo {pages} {4167} (\bibinfo {year} {2023})},\ \Eprint
      {https://arxiv.org/abs/2209.06299} {arXiv:2209.06299 [hep-ph]} \BibitemShut
      {NoStop}%
    \bibitem [{\citenamefont {Wadekar}\ and\ \citenamefont {Wang}(2022)}]{LeoT}%
      \BibitemOpen
      \bibfield  {author} {\bibinfo {author} {\bibfnamefont {D.}~\bibnamefont
      {Wadekar}}\ and\ \bibinfo {author} {\bibfnamefont {Z.}~\bibnamefont {Wang}},\
      }\href {https://doi.org/10.1103/PhysRevD.106.075007} {\bibfield  {journal}
      {\bibinfo  {journal} {Phys. Rev. D}\ }\textbf {\bibinfo {volume} {106}},\
      \bibinfo {pages} {075007} (\bibinfo {year} {2022})},\ \Eprint
      {https://arxiv.org/abs/2111.08025} {arXiv:2111.08025 [hep-ph]} \BibitemShut
      {NoStop}%
    \bibitem [{\citenamefont {Bernal}\ \emph {et~al.}(2023)\citenamefont {Bernal},
      \citenamefont {Caputo}, \citenamefont {Sato-Polito}, \citenamefont
      {Mirocha},\ and\ \citenamefont {Kamionkowski}}]{GRayAtt}%
      \BibitemOpen
      \bibfield  {author} {\bibinfo {author} {\bibfnamefont {J.~L.}\ \bibnamefont
      {Bernal}}, \bibinfo {author} {\bibfnamefont {A.}~\bibnamefont {Caputo}},
      \bibinfo {author} {\bibfnamefont {G.}~\bibnamefont {Sato-Polito}}, \bibinfo
      {author} {\bibfnamefont {J.}~\bibnamefont {Mirocha}},\ and\ \bibinfo {author}
      {\bibfnamefont {M.}~\bibnamefont {Kamionkowski}},\ }\href
      {https://doi.org/10.1103/PhysRevD.107.103046} {\bibfield  {journal} {\bibinfo
       {journal} {Phys. Rev. D}\ }\textbf {\bibinfo {volume} {107}},\ \bibinfo
      {pages} {103046} (\bibinfo {year} {2023})},\ \Eprint
      {https://arxiv.org/abs/2208.13794} {arXiv:2208.13794 [astro-ph.CO]}
      \BibitemShut {NoStop}%
    \bibitem [{\citenamefont {Carenza}\ \emph {et~al.}(2023)\citenamefont
      {Carenza}, \citenamefont {Lucente},\ and\ \citenamefont {Vitagliano}}]{HST}%
      \BibitemOpen
      \bibfield  {author} {\bibinfo {author} {\bibfnamefont {P.}~\bibnamefont
      {Carenza}}, \bibinfo {author} {\bibfnamefont {G.}~\bibnamefont {Lucente}},\
      and\ \bibinfo {author} {\bibfnamefont {E.}~\bibnamefont {Vitagliano}},\
      }\href {https://doi.org/10.1103/PhysRevD.107.083032} {\bibfield  {journal}
      {\bibinfo  {journal} {Phys. Rev. D}\ }\textbf {\bibinfo {volume} {107}},\
      \bibinfo {pages} {083032} (\bibinfo {year} {2023})},\ \Eprint
      {https://arxiv.org/abs/2301.06560} {arXiv:2301.06560 [hep-ph]} \BibitemShut
      {NoStop}%
    \bibitem [{\citenamefont {Nakayama}\ and\ \citenamefont {Yin}(2022)}]{HSTNak}%
      \BibitemOpen
      \bibfield  {author} {\bibinfo {author} {\bibfnamefont {K.}~\bibnamefont
      {Nakayama}}\ and\ \bibinfo {author} {\bibfnamefont {W.}~\bibnamefont {Yin}},\
      }\href {https://doi.org/10.1103/PhysRevD.106.103505} {\bibfield  {journal}
      {\bibinfo  {journal} {Phys. Rev. D}\ }\textbf {\bibinfo {volume} {106}},\
      \bibinfo {pages} {103505} (\bibinfo {year} {2022})},\ \Eprint
      {https://arxiv.org/abs/2205.01079} {arXiv:2205.01079 [hep-ph]} \BibitemShut
      {NoStop}%
    \bibitem [{\citenamefont {Grin}\ \emph {et~al.}(2007)\citenamefont {Grin},
      \citenamefont {Covone}, \citenamefont {Kneib}, \citenamefont {Kamionkowski},
      \citenamefont {Blain},\ and\ \citenamefont {Jullo}}]{VIMOS}%
      \BibitemOpen
      \bibfield  {author} {\bibinfo {author} {\bibfnamefont {D.}~\bibnamefont
      {Grin}}, \bibinfo {author} {\bibfnamefont {G.}~\bibnamefont {Covone}},
      \bibinfo {author} {\bibfnamefont {J.-P.}\ \bibnamefont {Kneib}}, \bibinfo
      {author} {\bibfnamefont {M.}~\bibnamefont {Kamionkowski}}, \bibinfo {author}
      {\bibfnamefont {A.}~\bibnamefont {Blain}},\ and\ \bibinfo {author}
      {\bibfnamefont {E.}~\bibnamefont {Jullo}},\ }\href
      {https://doi.org/10.1103/PhysRevD.75.105018} {\bibfield  {journal} {\bibinfo
      {journal} {Phys. Rev. D}\ }\textbf {\bibinfo {volume} {75}},\ \bibinfo
      {pages} {105018} (\bibinfo {year} {2007})},\ \Eprint
      {https://arxiv.org/abs/astro-ph/0611502} {arXiv:astro-ph/0611502}
      \BibitemShut {NoStop}%
    \bibitem [{\citenamefont {Liu}\ \emph {et~al.}(2023)\citenamefont {Liu},
      \citenamefont {Qin}, \citenamefont {Ridgway},\ and\ \citenamefont
      {Slatyer}}]{CMB1}%
      \BibitemOpen
      \bibfield  {author} {\bibinfo {author} {\bibfnamefont {H.}~\bibnamefont
      {Liu}}, \bibinfo {author} {\bibfnamefont {W.}~\bibnamefont {Qin}}, \bibinfo
      {author} {\bibfnamefont {G.~W.}\ \bibnamefont {Ridgway}},\ and\ \bibinfo
      {author} {\bibfnamefont {T.~R.}\ \bibnamefont {Slatyer}},\ }\href
      {https://doi.org/10.1103/PhysRevD.108.043531} {\bibfield  {journal} {\bibinfo
       {journal} {Phys. Rev. D}\ }\textbf {\bibinfo {volume} {108}},\ \bibinfo
      {pages} {043531} (\bibinfo {year} {2023})},\ \Eprint
      {https://arxiv.org/abs/2303.07370} {arXiv:2303.07370 [astro-ph.CO]}
      \BibitemShut {NoStop}%
    \bibitem [{\citenamefont {Capozzi}\ \emph {et~al.}(2023)\citenamefont
      {Capozzi}, \citenamefont {Ferreira}, \citenamefont {Lopez-Honorez},\ and\
      \citenamefont {Mena}}]{CMB2}%
      \BibitemOpen
      \bibfield  {author} {\bibinfo {author} {\bibfnamefont {F.}~\bibnamefont
      {Capozzi}}, \bibinfo {author} {\bibfnamefont {R.~Z.}\ \bibnamefont
      {Ferreira}}, \bibinfo {author} {\bibfnamefont {L.}~\bibnamefont
      {Lopez-Honorez}},\ and\ \bibinfo {author} {\bibfnamefont {O.}~\bibnamefont
      {Mena}},\ }\href {https://doi.org/10.1088/1475-7516/2023/06/060} {\bibfield
      {journal} {\bibinfo  {journal} {JCAP}\ }\textbf {\bibinfo {volume} {06}},\
      \bibinfo {pages} {060}},\ \Eprint {https://arxiv.org/abs/2303.07426}
      {arXiv:2303.07426 [astro-ph.CO]} \BibitemShut {NoStop}%
    \bibitem [{\citenamefont {Todarello}\ \emph {et~al.}(2023)\citenamefont
      {Todarello}, \citenamefont {Regis}, \citenamefont {Reynoso-Cordova},
      \citenamefont {Taoso}, \citenamefont {Vaz}, \citenamefont {Brinchmann},
      \citenamefont {Steinmetz},\ and\ \citenamefont {Zoutendijke}}]{MUSE1}%
      \BibitemOpen
      \bibfield  {author} {\bibinfo {author} {\bibfnamefont {E.}~\bibnamefont
      {Todarello}}, \bibinfo {author} {\bibfnamefont {M.}~\bibnamefont {Regis}},
      \bibinfo {author} {\bibfnamefont {J.}~\bibnamefont {Reynoso-Cordova}},
      \bibinfo {author} {\bibfnamefont {M.}~\bibnamefont {Taoso}}, \bibinfo
      {author} {\bibfnamefont {D.}~\bibnamefont {Vaz}}, \bibinfo {author}
      {\bibfnamefont {J.}~\bibnamefont {Brinchmann}}, \bibinfo {author}
      {\bibfnamefont {M.}~\bibnamefont {Steinmetz}},\ and\ \bibinfo {author}
      {\bibfnamefont {S.~L.}\ \bibnamefont {Zoutendijke}},\ }\href@noop {}
      {\bibinfo {title} {{Robust bounds on ALP dark matter from dwarf spheroidal
      galaxies in the optical MUSE-Faint survey}}} (\bibinfo {year} {2023}),\
      \Eprint {https://arxiv.org/abs/2307.07403} {arXiv:2307.07403 [astro-ph.CO]}
      \BibitemShut {NoStop}%
    \bibitem [{\citenamefont {Regis}\ \emph {et~al.}(2021)\citenamefont {Regis},
      \citenamefont {Taoso}, \citenamefont {Vaz}, \citenamefont {Brinchmann},
      \citenamefont {Zoutendijk}, \citenamefont {Bouch\'e},\ and\ \citenamefont
      {Steinmetz}}]{MUSE2}%
      \BibitemOpen
      \bibfield  {author} {\bibinfo {author} {\bibfnamefont {M.}~\bibnamefont
      {Regis}}, \bibinfo {author} {\bibfnamefont {M.}~\bibnamefont {Taoso}},
      \bibinfo {author} {\bibfnamefont {D.}~\bibnamefont {Vaz}}, \bibinfo {author}
      {\bibfnamefont {J.}~\bibnamefont {Brinchmann}}, \bibinfo {author}
      {\bibfnamefont {S.~L.}\ \bibnamefont {Zoutendijk}}, \bibinfo {author}
      {\bibfnamefont {N.~F.}\ \bibnamefont {Bouch\'e}},\ and\ \bibinfo {author}
      {\bibfnamefont {M.}~\bibnamefont {Steinmetz}},\ }\href
      {https://doi.org/10.1016/j.physletb.2021.136075} {\bibfield  {journal}
      {\bibinfo  {journal} {Phys. Lett. B}\ }\textbf {\bibinfo {volume} {814}},\
      \bibinfo {pages} {136075} (\bibinfo {year} {2021})},\ \Eprint
      {https://arxiv.org/abs/2009.01310} {arXiv:2009.01310 [astro-ph.CO]}
      \BibitemShut {NoStop}%
    \bibitem [{\citenamefont {Janish}\ and\ \citenamefont {Pinetti}(2023)}]{JWST}%
      \BibitemOpen
      \bibfield  {author} {\bibinfo {author} {\bibfnamefont {R.}~\bibnamefont
      {Janish}}\ and\ \bibinfo {author} {\bibfnamefont {E.}~\bibnamefont
      {Pinetti}},\ }\href@noop {} {\bibinfo {title} {{Hunting dark matter lines in
      the infrared background with the James Webb Space Telescope}}} (\bibinfo
      {year} {2023}),\ \Eprint {https://arxiv.org/abs/2310.15395} {arXiv:2310.15395
      [hep-ph]} \BibitemShut {NoStop}%
    \bibitem [{\citenamefont {Fong}\ \emph {et~al.}(2024)\citenamefont {Fong},
      \citenamefont {Ng},\ and\ \citenamefont {Liu}}]{ROSITA}%
      \BibitemOpen
      \bibfield  {author} {\bibinfo {author} {\bibfnamefont {C.}~\bibnamefont
      {Fong}}, \bibinfo {author} {\bibfnamefont {K.~C.~Y.}\ \bibnamefont {Ng}},\
      and\ \bibinfo {author} {\bibfnamefont {Q.}~\bibnamefont {Liu}},\ }\href
      {https://doi.org/10.22323/1.441.0172} {\bibfield  {journal} {\bibinfo
      {journal} {PoS}\ }\textbf {\bibinfo {volume} {TAUP2023}},\ \bibinfo {pages}
      {172} (\bibinfo {year} {2024})},\ \Eprint {https://arxiv.org/abs/2401.16747}
      {arXiv:2401.16747 [hep-ph]} \BibitemShut {NoStop}%
    \bibitem [{\citenamefont {Anastassopoulos}\ \emph {et~al.}(2017)\citenamefont
      {Anastassopoulos} \emph {et~al.}}]{CAST}%
      \BibitemOpen
      \bibfield  {author} {\bibinfo {author} {\bibfnamefont {V.}~\bibnamefont
      {Anastassopoulos}} \emph {et~al.} (\bibinfo {collaboration} {CAST}),\ }\href
      {https://doi.org/10.1038/nphys4109} {\bibfield  {journal} {\bibinfo
      {journal} {Nature Phys.}\ }\textbf {\bibinfo {volume} {13}},\ \bibinfo
      {pages} {584} (\bibinfo {year} {2017})},\ \Eprint
      {https://arxiv.org/abs/1705.02290} {arXiv:1705.02290 [hep-ex]} \BibitemShut
      {NoStop}%
    \bibitem [{\citenamefont {Flambaum}\ \emph {et~al.}(2022)\citenamefont
      {Flambaum}, \citenamefont {McAllister}, \citenamefont {Samsonov},\ and\
      \citenamefont {Tobar}}]{Tobar}%
      \BibitemOpen
      \bibfield  {author} {\bibinfo {author} {\bibfnamefont {V.~V.}\ \bibnamefont
      {Flambaum}}, \bibinfo {author} {\bibfnamefont {B.~T.}\ \bibnamefont
      {McAllister}}, \bibinfo {author} {\bibfnamefont {I.~B.}\ \bibnamefont
      {Samsonov}},\ and\ \bibinfo {author} {\bibfnamefont {M.~E.}\ \bibnamefont
      {Tobar}},\ }\href {https://doi.org/10.1103/PhysRevD.106.055037} {\bibfield
      {journal} {\bibinfo  {journal} {Phys. Rev. D}\ }\textbf {\bibinfo {volume}
      {106}},\ \bibinfo {pages} {055037} (\bibinfo {year} {2022})},\ \Eprint
      {https://arxiv.org/abs/2207.14437} {arXiv:2207.14437 [hep-ph]} \BibitemShut
      {NoStop}%
    \bibitem [{\citenamefont {Bottaro}\ \emph {et~al.}(2023)\citenamefont
      {Bottaro}, \citenamefont {Caputo}, \citenamefont {Raffelt},\ and\
      \citenamefont {Vitagliano}}]{Vitagliano}%
      \BibitemOpen
      \bibfield  {author} {\bibinfo {author} {\bibfnamefont {S.}~\bibnamefont
      {Bottaro}}, \bibinfo {author} {\bibfnamefont {A.}~\bibnamefont {Caputo}},
      \bibinfo {author} {\bibfnamefont {G.}~\bibnamefont {Raffelt}},\ and\ \bibinfo
      {author} {\bibfnamefont {E.}~\bibnamefont {Vitagliano}},\ }\href
      {https://doi.org/10.1088/1475-7516/2023/07/071} {\bibfield  {journal}
      {\bibinfo  {journal} {JCAP}\ }\textbf {\bibinfo {volume} {07}},\ \bibinfo
      {pages} {071}},\ \Eprint {https://arxiv.org/abs/2303.00778} {arXiv:2303.00778
      [hep-ph]} \BibitemShut {NoStop}%
    \bibitem [{\citenamefont {Chen}\ \emph {et~al.}(2017)\citenamefont {Chen},
      \citenamefont {Pospelov},\ and\ \citenamefont {Zhong}}]{MuonPospelov}%
      \BibitemOpen
      \bibfield  {author} {\bibinfo {author} {\bibfnamefont {C.-Y.}\ \bibnamefont
      {Chen}}, \bibinfo {author} {\bibfnamefont {M.}~\bibnamefont {Pospelov}},\
      and\ \bibinfo {author} {\bibfnamefont {Y.-M.}\ \bibnamefont {Zhong}},\ }\href
      {https://doi.org/10.1103/PhysRevD.95.115005} {\bibfield  {journal} {\bibinfo
      {journal} {Phys. Rev. D}\ }\textbf {\bibinfo {volume} {95}},\ \bibinfo
      {pages} {115005} (\bibinfo {year} {2017})},\ \Eprint
      {https://arxiv.org/abs/1701.07437} {arXiv:1701.07437 [hep-ph]} \BibitemShut
      {NoStop}%
    \bibitem [{\citenamefont {Jean}\ \emph {et~al.}(2003)\citenamefont {Jean} \emph
      {et~al.}}]{SPI-INTEGRAL}%
      \BibitemOpen
      \bibfield  {author} {\bibinfo {author} {\bibfnamefont {P.}~\bibnamefont
      {Jean}} \emph {et~al.},\ }\href {https://doi.org/10.1051/0004-6361:20031056}
      {\bibfield  {journal} {\bibinfo  {journal} {Astron. Astrophys.}\ }\textbf
      {\bibinfo {volume} {407}},\ \bibinfo {pages} {L55} (\bibinfo {year}
      {2003})},\ \Eprint {https://arxiv.org/abs/astro-ph/0309484}
      {arXiv:astro-ph/0309484} \BibitemShut {NoStop}%
    \bibitem [{\citenamefont {Flambaum}\ and\ \citenamefont
      {Samsonov}(2021)}]{QN1}%
      \BibitemOpen
      \bibfield  {author} {\bibinfo {author} {\bibfnamefont {V.~V.}\ \bibnamefont
      {Flambaum}}\ and\ \bibinfo {author} {\bibfnamefont {I.~B.}\ \bibnamefont
      {Samsonov}},\ }\href {https://doi.org/10.1103/PhysRevD.104.063042} {\bibfield
       {journal} {\bibinfo  {journal} {Phys. Rev. D}\ }\textbf {\bibinfo {volume}
      {104}},\ \bibinfo {pages} {063042} (\bibinfo {year} {2021})},\ \Eprint
      {https://arxiv.org/abs/2108.00652} {arXiv:2108.00652 [hep-ph]} \BibitemShut
      {NoStop}%
    \bibitem [{\citenamefont {Stadnik}(2023)}]{StadnikMuon}%
      \BibitemOpen
      \bibfield  {author} {\bibinfo {author} {\bibfnamefont {Y.~V.}\ \bibnamefont
      {Stadnik}},\ }\href {https://doi.org/10.1103/PhysRevLett.131.011001}
      {\bibfield  {journal} {\bibinfo  {journal} {Phys. Rev. Lett.}\ }\textbf
      {\bibinfo {volume} {131}},\ \bibinfo {pages} {011001} (\bibinfo {year}
      {2023})},\ \Eprint {https://arxiv.org/abs/2206.10808} {arXiv:2206.10808
      [hep-ph]} \BibitemShut {NoStop}%
    \bibitem [{\citenamefont {Caputo}\ \emph {et~al.}(2022)\citenamefont {Caputo},
      \citenamefont {Raffelt},\ and\ \citenamefont {Vitagliano}}]{Raffelt}%
      \BibitemOpen
      \bibfield  {author} {\bibinfo {author} {\bibfnamefont {A.}~\bibnamefont
      {Caputo}}, \bibinfo {author} {\bibfnamefont {G.}~\bibnamefont {Raffelt}},\
      and\ \bibinfo {author} {\bibfnamefont {E.}~\bibnamefont {Vitagliano}},\
      }\href {https://doi.org/10.1103/PhysRevD.105.035022} {\bibfield  {journal}
      {\bibinfo  {journal} {Phys. Rev. D}\ }\textbf {\bibinfo {volume} {105}},\
      \bibinfo {pages} {035022} (\bibinfo {year} {2022})},\ \Eprint
      {https://arxiv.org/abs/2109.03244} {arXiv:2109.03244 [hep-ph]} \BibitemShut
      {NoStop}%
    \bibitem [{\citenamefont {Leutwyler}\ and\ \citenamefont
      {Shifman}(1989)}]{Leutwyler89}%
      \BibitemOpen
      \bibfield  {author} {\bibinfo {author} {\bibfnamefont {H.}~\bibnamefont
      {Leutwyler}}\ and\ \bibinfo {author} {\bibfnamefont {M.~A.}\ \bibnamefont
      {Shifman}},\ }\href {https://doi.org/10.1016/0370-2693(89)91730-9} {\bibfield
       {journal} {\bibinfo  {journal} {Phys. Lett. B}\ }\textbf {\bibinfo {volume}
      {221}},\ \bibinfo {pages} {384} (\bibinfo {year} {1989})}\BibitemShut
      {NoStop}%
    \bibitem [{\citenamefont {Lee}\ \emph {et~al.}(2020)\citenamefont {Lee},
      \citenamefont {Adelberger}, \citenamefont {Cook}, \citenamefont {Fleischer},\
      and\ \citenamefont {Heckel}}]{EP1}%
      \BibitemOpen
      \bibfield  {author} {\bibinfo {author} {\bibfnamefont {J.~G.}\ \bibnamefont
      {Lee}}, \bibinfo {author} {\bibfnamefont {E.~G.}\ \bibnamefont {Adelberger}},
      \bibinfo {author} {\bibfnamefont {T.~S.}\ \bibnamefont {Cook}}, \bibinfo
      {author} {\bibfnamefont {S.~M.}\ \bibnamefont {Fleischer}},\ and\ \bibinfo
      {author} {\bibfnamefont {B.~R.}\ \bibnamefont {Heckel}},\ }\href
      {https://doi.org/10.1103/PhysRevLett.124.101101} {\bibfield  {journal}
      {\bibinfo  {journal} {Phys. Rev. Lett.}\ }\textbf {\bibinfo {volume} {124}},\
      \bibinfo {pages} {101101} (\bibinfo {year} {2020})},\ \Eprint
      {https://arxiv.org/abs/2002.11761} {arXiv:2002.11761 [hep-ex]} \BibitemShut
      {NoStop}%
    \bibitem [{\citenamefont {Kapner}\ \emph {et~al.}(2007)\citenamefont {Kapner},
      \citenamefont {Cook}, \citenamefont {Adelberger}, \citenamefont {Gundlach},
      \citenamefont {Heckel}, \citenamefont {Hoyle},\ and\ \citenamefont
      {Swanson}}]{EP2}%
      \BibitemOpen
      \bibfield  {author} {\bibinfo {author} {\bibfnamefont {D.~J.}\ \bibnamefont
      {Kapner}}, \bibinfo {author} {\bibfnamefont {T.~S.}\ \bibnamefont {Cook}},
      \bibinfo {author} {\bibfnamefont {E.~G.}\ \bibnamefont {Adelberger}},
      \bibinfo {author} {\bibfnamefont {J.~H.}\ \bibnamefont {Gundlach}}, \bibinfo
      {author} {\bibfnamefont {B.~R.}\ \bibnamefont {Heckel}}, \bibinfo {author}
      {\bibfnamefont {C.~D.}\ \bibnamefont {Hoyle}},\ and\ \bibinfo {author}
      {\bibfnamefont {H.~E.}\ \bibnamefont {Swanson}},\ }\href
      {https://doi.org/10.1103/PhysRevLett.98.021101} {\bibfield  {journal}
      {\bibinfo  {journal} {Phys. Rev. Lett.}\ }\textbf {\bibinfo {volume} {98}},\
      \bibinfo {pages} {021101} (\bibinfo {year} {2007})},\ \Eprint
      {https://arxiv.org/abs/hep-ph/0611184} {arXiv:hep-ph/0611184} \BibitemShut
      {NoStop}%
    \bibitem [{\citenamefont {Adelberger}\ \emph {et~al.}(2007)\citenamefont
      {Adelberger}, \citenamefont {Heckel}, \citenamefont {Hoedl}, \citenamefont
      {Hoyle}, \citenamefont {Kapner},\ and\ \citenamefont {Upadhye}}]{EP3}%
      \BibitemOpen
      \bibfield  {author} {\bibinfo {author} {\bibfnamefont {E.~G.}\ \bibnamefont
      {Adelberger}}, \bibinfo {author} {\bibfnamefont {B.~R.}\ \bibnamefont
      {Heckel}}, \bibinfo {author} {\bibfnamefont {S.}~\bibnamefont {Hoedl}},
      \bibinfo {author} {\bibfnamefont {C.~D.}\ \bibnamefont {Hoyle}}, \bibinfo
      {author} {\bibfnamefont {D.~J.}\ \bibnamefont {Kapner}},\ and\ \bibinfo
      {author} {\bibfnamefont {A.}~\bibnamefont {Upadhye}},\ }\href
      {https://doi.org/10.1103/PhysRevLett.98.131104} {\bibfield  {journal}
      {\bibinfo  {journal} {Phys. Rev. Lett.}\ }\textbf {\bibinfo {volume} {98}},\
      \bibinfo {pages} {131104} (\bibinfo {year} {2007})},\ \Eprint
      {https://arxiv.org/abs/hep-ph/0611223} {arXiv:hep-ph/0611223} \BibitemShut
      {NoStop}%
    \bibitem [{\citenamefont {Hoyle}\ \emph {et~al.}(2004)\citenamefont {Hoyle},
      \citenamefont {Kapner}, \citenamefont {Heckel}, \citenamefont {Adelberger},
      \citenamefont {Gundlach}, \citenamefont {Schmidt},\ and\ \citenamefont
      {Swanson}}]{EP4}%
      \BibitemOpen
      \bibfield  {author} {\bibinfo {author} {\bibfnamefont {C.~D.}\ \bibnamefont
      {Hoyle}}, \bibinfo {author} {\bibfnamefont {D.~J.}\ \bibnamefont {Kapner}},
      \bibinfo {author} {\bibfnamefont {B.~R.}\ \bibnamefont {Heckel}}, \bibinfo
      {author} {\bibfnamefont {E.~G.}\ \bibnamefont {Adelberger}}, \bibinfo
      {author} {\bibfnamefont {J.~H.}\ \bibnamefont {Gundlach}}, \bibinfo {author}
      {\bibfnamefont {U.}~\bibnamefont {Schmidt}},\ and\ \bibinfo {author}
      {\bibfnamefont {H.~E.}\ \bibnamefont {Swanson}},\ }\href
      {https://doi.org/10.1103/PhysRevD.70.042004} {\bibfield  {journal} {\bibinfo
      {journal} {Phys. Rev. D}\ }\textbf {\bibinfo {volume} {70}},\ \bibinfo
      {pages} {042004} (\bibinfo {year} {2004})},\ \Eprint
      {https://arxiv.org/abs/hep-ph/0405262} {arXiv:hep-ph/0405262} \BibitemShut
      {NoStop}%
    \bibitem [{\citenamefont {Hoyle}\ \emph {et~al.}(2001)\citenamefont {Hoyle},
      \citenamefont {Schmidt}, \citenamefont {Heckel}, \citenamefont {Adelberger},
      \citenamefont {Gundlach}, \citenamefont {Kapner},\ and\ \citenamefont
      {Swanson}}]{EP5}%
      \BibitemOpen
      \bibfield  {author} {\bibinfo {author} {\bibfnamefont {C.~D.}\ \bibnamefont
      {Hoyle}}, \bibinfo {author} {\bibfnamefont {U.}~\bibnamefont {Schmidt}},
      \bibinfo {author} {\bibfnamefont {B.~R.}\ \bibnamefont {Heckel}}, \bibinfo
      {author} {\bibfnamefont {E.~G.}\ \bibnamefont {Adelberger}}, \bibinfo
      {author} {\bibfnamefont {J.~H.}\ \bibnamefont {Gundlach}}, \bibinfo {author}
      {\bibfnamefont {D.~J.}\ \bibnamefont {Kapner}},\ and\ \bibinfo {author}
      {\bibfnamefont {H.~E.}\ \bibnamefont {Swanson}},\ }\href
      {https://doi.org/10.1103/PhysRevLett.86.1418} {\bibfield  {journal} {\bibinfo
       {journal} {Phys. Rev. Lett.}\ }\textbf {\bibinfo {volume} {86}},\ \bibinfo
      {pages} {1418} (\bibinfo {year} {2001})},\ \Eprint
      {https://arxiv.org/abs/hep-ph/0011014} {arXiv:hep-ph/0011014} \BibitemShut
      {NoStop}%
    \bibitem [{\citenamefont {Smith}\ \emph {et~al.}(2000)\citenamefont {Smith},
      \citenamefont {Hoyle}, \citenamefont {Gundlach}, \citenamefont {Adelberger},
      \citenamefont {Heckel},\ and\ \citenamefont {Swanson}}]{EP6}%
      \BibitemOpen
      \bibfield  {author} {\bibinfo {author} {\bibfnamefont {G.~L.}\ \bibnamefont
      {Smith}}, \bibinfo {author} {\bibfnamefont {C.~D.}\ \bibnamefont {Hoyle}},
      \bibinfo {author} {\bibfnamefont {J.~H.}\ \bibnamefont {Gundlach}}, \bibinfo
      {author} {\bibfnamefont {E.~G.}\ \bibnamefont {Adelberger}}, \bibinfo
      {author} {\bibfnamefont {B.~R.}\ \bibnamefont {Heckel}},\ and\ \bibinfo
      {author} {\bibfnamefont {H.~E.}\ \bibnamefont {Swanson}},\ }\href
      {https://doi.org/10.1103/PhysRevD.61.022001} {\bibfield  {journal} {\bibinfo
      {journal} {Phys. Rev. D}\ }\textbf {\bibinfo {volume} {61}},\ \bibinfo
      {pages} {022001} (\bibinfo {year} {2000})}\BibitemShut {NoStop}%
    \bibitem [{\citenamefont {Leefer}\ \emph {et~al.}(2016)\citenamefont {Leefer},
      \citenamefont {Gerhardus}, \citenamefont {Budker}, \citenamefont {Flambaum},\
      and\ \citenamefont {Stadnik}}]{Leefer2016}%
      \BibitemOpen
      \bibfield  {author} {\bibinfo {author} {\bibfnamefont {N.}~\bibnamefont
      {Leefer}}, \bibinfo {author} {\bibfnamefont {A.}~\bibnamefont {Gerhardus}},
      \bibinfo {author} {\bibfnamefont {D.}~\bibnamefont {Budker}}, \bibinfo
      {author} {\bibfnamefont {V.~V.}\ \bibnamefont {Flambaum}},\ and\ \bibinfo
      {author} {\bibfnamefont {Y.~V.}\ \bibnamefont {Stadnik}},\ }\href
      {https://doi.org/10.1103/PhysRevLett.117.271601} {\bibfield  {journal}
      {\bibinfo  {journal} {Phys. Rev. Lett.}\ }\textbf {\bibinfo {volume} {117}},\
      \bibinfo {pages} {271601} (\bibinfo {year} {2016})},\ \Eprint
      {https://arxiv.org/abs/1607.04956} {arXiv:1607.04956 [physics.atom-ph]}
      \BibitemShut {NoStop}%
    \bibitem [{\citenamefont {Dzuba}\ \emph {et~al.}(2023)\citenamefont {Dzuba},
      \citenamefont {Flambaum},\ and\ \citenamefont {Samsonov}}]{Dzuba23}%
      \BibitemOpen
      \bibfield  {author} {\bibinfo {author} {\bibfnamefont {V.~A.}\ \bibnamefont
      {Dzuba}}, \bibinfo {author} {\bibfnamefont {V.~V.}\ \bibnamefont
      {Flambaum}},\ and\ \bibinfo {author} {\bibfnamefont {I.~B.}\ \bibnamefont
      {Samsonov}},\ }\href{https://doi.org/10.1103/PhysRevD.109.115032} {\bibinfo {journal} {Phys. Rev. D}\ \textbf {\bibinfo {volume}{109}},\ \bibinfo {pages} {115032} (\bibinfo {year} {2024}),}\ \Eprint
      {https://arxiv.org/abs/2312.10566} {arXiv:2312.10566 [hep-ph]} \BibitemShut
      {NoStop}%
    \bibitem [{\citenamefont {Piazza}\ and\ \citenamefont
      {Pospelov}(2010)}]{PospelovHiggs}%
      \BibitemOpen
      \bibfield  {author} {\bibinfo {author} {\bibfnamefont {F.}~\bibnamefont
      {Piazza}}\ and\ \bibinfo {author} {\bibfnamefont {M.}~\bibnamefont
      {Pospelov}},\ }\href {https://doi.org/10.1103/PhysRevD.82.043533} {\bibfield
      {journal} {\bibinfo  {journal} {Phys. Rev. D}\ }\textbf {\bibinfo {volume}
      {82}},\ \bibinfo {pages} {043533} (\bibinfo {year} {2010})},\ \Eprint
      {https://arxiv.org/abs/1003.2313} {arXiv:1003.2313 [hep-ph]} \BibitemShut
      {NoStop}%
    \bibitem [{\citenamefont {Stadnik}\ and\ \citenamefont
      {Flambaum}(2016)}]{StadnikHiggs}%
      \BibitemOpen
      \bibfield  {author} {\bibinfo {author} {\bibfnamefont {Y.~V.}\ \bibnamefont
      {Stadnik}}\ and\ \bibinfo {author} {\bibfnamefont {V.~V.}\ \bibnamefont
      {Flambaum}},\ }\href {https://doi.org/10.1103/PhysRevA.94.022111} {\bibfield
      {journal} {\bibinfo  {journal} {Phys. Rev. A}\ }\textbf {\bibinfo {volume}
      {94}},\ \bibinfo {pages} {022111} (\bibinfo {year} {2016})},\ \Eprint
      {https://arxiv.org/abs/1605.04028} {arXiv:1605.04028 [physics.atom-ph]}
      \BibitemShut {NoStop}%
    \bibitem [{\citenamefont {Workman}\ \emph {et~al.}(2022)\citenamefont {Workman}
      \emph {et~al.}}]{PDG}%
      \BibitemOpen
      \bibfield  {author} {\bibinfo {author} {\bibfnamefont {R.~L.}\ \bibnamefont
      {Workman}} \emph {et~al.},\ }\href {https://doi.org/10.1093/ptep/ptac097}
      {\bibfield  {journal} {\bibinfo  {journal} {Prog. Theor. Exp. Phys.}\
      }\textbf {\bibinfo {volume} {2022}},\ \bibinfo {pages} {083C01} (\bibinfo
      {year} {2022})}\BibitemShut {NoStop}%
    \end{thebibliography}
\end{document}